\documentclass[twocolumn]{aastex61}
\usepackage{amsmath,amstext}
\usepackage[T1]{fontenc}
\usepackage{apjfonts} 
\usepackage[figure,figure*]{hypcap}
\usepackage{url}

\newcommand{\Msun}{\ensuremath{M_\odot}}

\defcitealias{curtin2017hscspec}{C19}

\shorttitle{First SHIZUCA results. I. Photometry}
\shortauthors{T. J. Moriya et al.}

\received{January 23, 2018}
\revised{February 10, 2019}
\accepted{February 12, 2019}
\journalinfo{Astrophysical Journal Supplements}

\begin{document}

\title{
First release of high-redshift superluminous supernovae from \\
the Subaru high-z supernova campaign (SHIZUCA). \\
I. Photometric properties.
}


\author[0000-0003-1169-1954]{Takashi J. Moriya}
\affiliation{National Astronomical Observatory of Japan, National Institutes of Natural Sciences, 2-21-1 Osawa, Mitaka, Tokyo 181-8588, Japan}

\author{Masaomi Tanaka}
\affiliation{Astronomical Institute, Tohoku University, 6-3 Aramaki Aza-Aoba, Aoba-ku, Sendai 980-8578, Japan}
\author{Naoki Yasuda}
\affiliation{Kavli Institute for the Physics and Mathematics of the Universe (WPI), The University of Tokyo Institutes for Advanced Study, The University of Tokyo, 5-1-5 Kashiwanoha, Kashiwa, Chiba 277-8583, Japan}

\author{Ji-an Jiang}
\affiliation{Institute of Astronomy, Graduate School of Science, The University of Tokyo, 2-21-1 Osawa, Mitaka, Tokyo 181-0015, Japan}
\author{Chien-Hsiu Lee}
\affiliation{Subaru Telescope, NAOJ, 650 N Aohoku Pl., Hilo, HI 96720, USA}
\author{Keiichi Maeda}
\affiliation{Department of Astronomy, Kyoto University, Kitashirakawa-Oiwake-cho, Sakyo-ku, Kyoto 606-8502, Japan}
\affiliation{Kavli Institute for the Physics and Mathematics of the Universe (WPI), The University of Tokyo Institutes for Advanced Study, The University of Tokyo, 5-1-5 Kashiwanoha, Kashiwa, Chiba 277-8583, Japan}
\author{Tomoki Morokuma}
\affiliation{Institute of Astronomy, Graduate School of Science, The University of Tokyo, 2-21-1 Osawa, Mitaka, Tokyo 181-0015, Japan}
\author{Ken'ichi Nomoto}
\affiliation{Kavli Institute for the Physics and Mathematics of the Universe (WPI), The University of Tokyo Institutes for Advanced Study, The University of Tokyo, 5-1-5 Kashiwanoha, Kashiwa, Chiba 277-8583, Japan}
\author{Robert M. Quimby}
\affiliation{Department of Astronomy / Mount Laguna Observatory, San Diego State University, 5500 Campanile Drive, San Diego, CA, 92812-1221, USA}
\affiliation{Kavli Institute for the Physics and Mathematics of the Universe (WPI), The University of Tokyo Institutes for Advanced Study, The University of Tokyo, 5-1-5 Kashiwanoha, Kashiwa, Chiba 277-8583, Japan}
\author{Nao Suzuki}
\affiliation{Kavli Institute for the Physics and Mathematics of the Universe (WPI), The University of Tokyo Institutes for Advanced Study, The University of Tokyo, 5-1-5 Kashiwanoha, Kashiwa, Chiba 277-8583, Japan}
\author{Ichiro Takahashi}
\affiliation{Kavli Institute for the Physics and Mathematics of the Universe (WPI), The University of Tokyo Institutes for Advanced Study, The University of Tokyo, 5-1-5 Kashiwanoha, Kashiwa, Chiba 277-8583, Japan}
\author{Masayuki Tanaka}
\affiliation{National Astronomical Observatory of Japan, National Institutes of Natural Sciences, 2-21-1 Osawa, Mitaka, Tokyo 181-8588, Japan}
\author{Nozomu Tominaga}
\affiliation{Department of Physics, Faculty of Science and Engineering, Konan University, 8-9-1 Okamoto, Kobe, Hyogo 658-8501, Japan}
\affiliation{Kavli Institute for the Physics and Mathematics of the Universe (WPI), The University of Tokyo Institutes for Advanced Study, The University of Tokyo, 5-1-5 Kashiwanoha, Kashiwa, Chiba 277-8583, Japan}
\author{Masaki Yamaguchi}
\affiliation{Institute of Astronomy, Graduate School of Science, The University of Tokyo, 2-21-1 Osawa, Mitaka, Tokyo 181-0015, Japan}

\author{Stephanie R. Bernard}
\affiliation{School of Physics, University of Melbourne, Parkville VIC 3010, Australia}
\affiliation{ARC Centre of Excellence for All-Sky Astrophysics (CAASTRO)}
\author{Jeff Cooke}
\affiliation{Centre for Astrophysics \& Supercomputing, Swinburne University of Technology, Hawthorn, VIC 3122, Australia}
\affiliation{ARC Centre of Excellence for All-Sky Astrophysics (CAASTRO)}
\author{Chris Curtin}
\affiliation{Centre for Astrophysics \& Supercomputing, Swinburne University of Technology, Hawthorn, VIC 3122, Australia}
\affiliation{ARC Centre of Excellence for All-Sky Astrophysics (CAASTRO)}
\author[0000-0002-1296-6887]{Llu\'is Galbany}
\affiliation{PITT PACC, Department of Physics and Astronomy, University of Pittsburgh, Pittsburgh, PA 15260, USA}
\author{Santiago Gonz\'alez-Gait\'an}
\affiliation{CENTRA, Instituto Superior T\'ecnico, Universidade de Lisboa, Portugal}
\author{Giuliano Pignata}
\affiliation{Departamento de Ciencias F\'isicas, Universidad Andres Bello, Avda. Rep\'ublica 252, Santiago, 8320000, Chile}
\affiliation{Millennium Institute of Astrophysics (MAS), Nuncio Monse\~nor S\'otero Sanz 100, Providencia, Santiago, Chile}
\author{Tyler Pritchard}
\affiliation{Centre for Astrophysics \& Supercomputing, Swinburne University of Technology, Hawthorn, VIC 3122, Australia}

\author{Yutaka Komiyama}
\affiliation{National Astronomical Observatory of Japan, National Institutes of Natural Sciences, 2-21-1 Osawa, Mitaka, Tokyo 181-8588, Japan}
\affiliation{Graduate University for Advanced Studies (SOKENDAI), 2-21-1 Osawa, Mitaka, Tokyo 181-8588, Japan}
\author{Robert H. Lupton}
\affiliation{Department of Astrophysical Sciences, Princeton University, 4 Ivy Lane, Princeton, NJ 08544, USA}



\begin{abstract}
We report our first discoveries of high-redshift supernovae from the Subaru HIgh-Z sUpernova CAmpaign (SHIZUCA), the transient survey using Subaru/Hyper Suprime-Cam. We report the discovery of three supernovae at the spectroscopically-confirmed redshifts of 2.399 (HSC16adga), 1.965 (HSC17auzg), and 1.851 (HSC17dbpf), and two supernova candidates with the host-galaxy photometric redshifts of 3.2 (HSC16apuo) and 4.2 (HSC17dsid), respectively. In this paper, we present their photometric properties and the spectroscopic properties of the confirmed high-redshift supernovae are presented in the accompanying paper \citep{curtin2017hscspec}. The supernovae with the confirmed redshifts of $z\simeq 2$ have the rest ultraviolet peak magnitudes close to $-21$~mag and they are likely superluminous supernovae. The discovery of three supernovae at $z\simeq 2$ roughly corresponds to the approximate event rate of $\sim 900\pm520~\mathrm{Gpc^{-3}~\mathrm{yr^{-1}}}$ with Poisson error, which is consistent with the total superluminous supernova rate estimated by extrapolating the local rate based on the cosmic star-formation history. Adding unconfirmed superluminous supernova candidates would increase the event rate. Our superluminous supernova candidates at the redshifts of around 3 and 4 indicate the approximate superluminous supernova rates of $\sim 400\pm400~\mathrm{Gpc^{-3}~yr^{-1}}$ ($z\sim 3$) and $\sim 500\pm500~\mathrm{Gpc^{-3}~yr^{-1}}$ ($z\sim4$) with Poisson errors.
Our initial results demonstrate the outstanding capability of Hyper Suprime-Cam to discover high-redshift supernovae.
\end{abstract}


\section{Introduction}
Supernovae (SNe) are luminous explosions of stars. Because of their huge luminosities, SNe can be observed even if they are far away, and they have indeed been used to explore the distant Universe. For example, Type~Ia SNe are known to be standardizable candles \citep[e.g.,][]{phillips2005relation}. The use of Type~Ia SNe to measure distances in the Universe led the discovery of the accelerating expansion of the Universe \citep{perlmutter1999acceleration,riess1998acceleration}.

Type~Ia SNe, which are explosions of white dwarfs, are usually brighter than core-collapse SNe that originate from massive stars \citep[e.g.,][]{richardson2014lumfunc}. Thus, Type~Ia SNe have long been used to probe the distant Universe \citep[e.g.,][]{suzuki2012cosmology}. However, recent transient surveys revealed the existence of core-collapse SNe that are much brighter and bluer than Type~Ia SNe. For example, Type~IIn SNe gain their luminosity through the interaction between the SN ejecta and circumstellar media \citep[e.g.,][]{moriya2013iinana} and, therefore, can be brighter and bluer than normal core-collapse SNe \citep[e.g.,][]{fransson2014sn2010jl}. Another example is so-called superluminous SNe (SLSNe) that can become brighter than $\sim -21~\mathrm{mag}$ and have blue spectra \citep[e.g.,][]{smith2010sn2006gyspectra,quimby2011slsn}. The mechanisms to make SLSNe very luminous are not yet fully understood, but their progenitors are believed to be massive stars \citep[e.g.,][]{gal-yam2012slsnrev,gal-yam2018slsnreview,howell2017slsnrev,moriya2018slsnreview}. At least some SLSNe are known to be Type~IIn \citep[e.g.,][]{smith2010sn2006gyspectra} and they are likely powered by the circumstellar interaction \citep[e.g.,][]{chevalier2011irwin,moriya2013sn2006gy,chatzopoulos2013anachi}. SLSNe that are not Type~IIn are mostly Type~I having broad carbon and oxygen features \citep[e.g.,][]{quimby2011slsn,howell2013slsn,yan2017gaia16apduv,mazzali2016slsn}. These core-collapse SNe can be detected well beyond the reach of Type~Ia SNe (e.g., \citealt{cooke2009highziindetec,cooke2012highzslsn,howell2013slsn,pan2017slsnz1p861,smith2017z2des}, but see also \citealt{rubin2017z2p22snia}) and enable us to study properties of massive stars in the early Universe. Some of them may even be used as standard candles \citep{blinnikov2012typiincosmology,quimby2013slsnrate,inserra2014slsncosmology,inserra2017euclid,scovacricchi2016slsncosmology}.

To find high-redshift SNe, it is necessary to conduct a deep and wide transient survey. Hyper Suprime-Cam (HSC, \citealt{miyazaki2017hsc,komiyama2017hsc,kawanomoto2017hsc,furusawa2017hsc}) on the 8.2~m Subaru telescope has a field-of-view of $1.8~\mathrm{deg^2}$ and it is one of the best instruments in the world to conduct such a wide and deep transient survey \citep[e.g.,][]{tanaka2016hscrapidrise}. To make use of its unique capability, a deep and wide transient survey was conducted with HSC from November 2016 to May 2017 under the HSC Subaru Strategic Program (SSP; Yasuda et al. in preparation, \citealt{aihara2017hscssp}). The HSC-SSP transient survey aims for detecting Type~Ia SNe at $1.0\lesssim z \lesssim 1.5$ with which a better constraint on the cosmological parameters can be obtained (Suzuki et al. in preparation). However, the same data can also be used to find luminous core-collapse SNe at $z\gtrsim 1.5$ \citep[e.g.,][]{cooke2008highziinpred,tanaka2012optslsndetec,tanaka2013nirslsndetec} and we conducted the Subaru HIgh-Z sUpernova CAmpaign (SHIZUCA). In this paper, we report our first discoveries of such high-redshift SNe beyond the redshift of 1.5 during the first half year of SHIZUCA. A similar half-year survey is planned in 2019--2020.

The rest of this paper is organized as follows.  First, we provide a brief summary of the half-year transient survey in Section~\ref{sec:serveyoverview}. We present high-redshift SNe at $z\simeq 2$ whose redshifts are confirmed by the spectroscopic follow-up observations with Keck/Low Resolution Imaging Spectrometer (LRIS) reported in the accompanying paper (\citealt{curtin2017hscspec}, \citetalias{curtin2017hscspec} hereafter) in Section~\ref{sec:specconfhighz}. Then, we report one SN candidate at $z\sim 3$ and another SN candidate at $z\sim 4$ in Section~\ref{sec:highzcand}. We are not able to obtain their spectra, but the photometric redshifts of their host galaxies 
suggest the high-redshift nature of the SNe. We discuss our results in Section~\ref{sec:discussion} and conclude this paper in Section~\ref{sec:conclusions}. The standard cosmology with $H_0=70~\mathrm{km~s^{-1}~Mpc^{-1}}$, $\Omega_\Lambda=0.7$, and $\Omega_M=0.3$ is adopted when necessary. All observed photometry is presented in the AB magnitude system.

\begin{deluxetable*}{lcccccccc}
\tablecaption{List of SNe and SN candidates. \label{tab:listofsne}}
\tablecolumns{4}
\tablenum{1}
\tablewidth{0pt}
\tablehead{
 &  &  & \multicolumn5c{host galaxy magnitudes in the HSC filters} & \\
\colhead{HSC name} & \colhead{IAU name} & \colhead{redshift} & \colhead{$g$} & \colhead{$r$} & \colhead{$i$} & \colhead{$z$} & \colhead{$y$} &\colhead{Section} 
}
\startdata
HSC16adga & SN~2016jhm & $2.399\pm 0.004$\tablenotemark{a}& $24.55\pm 0.03$& $24.42\pm 0.04$ & $24.48\pm 0.06$ & $24.29\pm 0.07$ & $24.20\pm 0.13$ & \ref{sec:hsc16adga} \\
HSC17auzg & SN~2016jhn & $1.965\pm 0.004$\tablenotemark{a}& $23.88\pm 0.02$ & $23.77 \pm 0.02$ & $23.54\pm0.02$ & $23.41\pm 0.03$ & $23.58\pm0.06$ &\ref{sec:hsc17auzg} \\
HSC17dbpf& SN~2017fei & $1.851\pm 0.004$\tablenotemark{a}& $24.11\pm 0.02$ & $23.91\pm 0.02$ & $23.67\pm 0.03$ & $23.63\pm 0.04$ & $23.60\pm0.08$ &\ref{sec:hsc17dbpf} \\
\hline
HSC16apuo & AT~2016jho & $2.82^{+0.47}_{-0.70}$\tablenotemark{b}& $27.00\pm 0.75$ & $25.31\pm 0.19$ & $25.50 \pm 0.35$ & $24.92\pm 0.29$ & $26.10\pm0.29$ & \ref{sec:hsc16apuo} \\
HSC17dsid& AT~2017fej & $4.20^{+0.09}_{-0.13}$\tablenotemark{b} & $27.74\pm 0.34$ & $25.07\pm 0.04$ & $24.83\pm 0.04$ & $24.68\pm 0.05$ & $25.23\pm 0.18$ & \ref{sec:hsc17dsid} \\
\enddata
\tablenotetext{a}{Spectroscopically confirmed (\citetalias{curtin2017hscspec}).}
\tablenotetext{b}{COSMOS2015 photometric redshift \citep{laigle2016cosmos2015}.}
\end{deluxetable*}

\section{Transient survey overview}\label{sec:serveyoverview}
We briefly summarize the transient survey conducted with Subaru/HSC from November 2016 to May 2017 as a part of the HSC-SSP survey \citep{aihara2017hscssp}. Details of the transient survey will be presented in Yasuda et al. (in preparation).

\begin{figure}
 \begin{center}
  \includegraphics[width=\columnwidth]{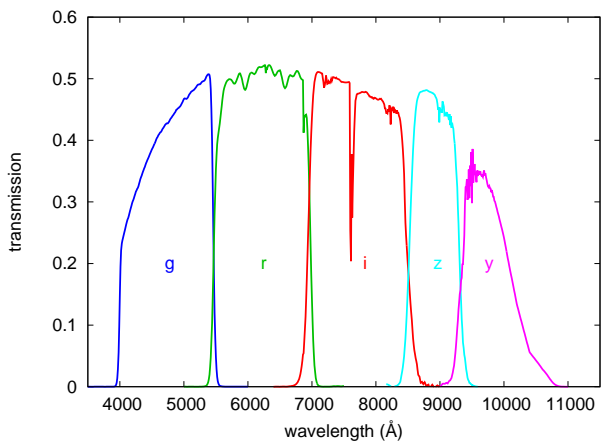}  
 \end{center}
\caption{
HSC filter transmissions, including the mirror reflectivity, transmissions of all the optics and filters, the atmosphere, and the response of the CCDs at the airmass of 1.2.
The wavelength ranges covered by the filters are $4000-5450$~\AA\ ($g$), $5450-7000$~\AA\ ($r$), $7000-8550$~\AA\ ($i$), $8550-9300$~\AA\ ($z$), and $9300-10700$~\AA\ ($y$).
}\label{fig:filters}
\end{figure}

The transient survey was performed in the HSC-SSP UltraDeep COSMOS field \citep{capak2007cosmos1strelease}.  The survey area is one field-of-view of HSC ($1.8~\mathrm{deg^2}$).  The observations with each filter (Fig.~\ref{fig:filters}) were performed for a few epochs at around every new moon. The data are reduced with hscPipe \citep{bosch2017hscpipe}, a version of the LSST stack \citep{ivezic2008lsst,axelrod2010lsst,juric2015lsst}.  The astrometry and photometry are calibrated relative to the Pan-STARRS1 (PS1) $3\pi$ catalog \citep{magnier2013ps1photo,schlafly2012ps1photo,tonry2012ps1photo}. The final photometry is obtained via point-spread function photometry in the template-subtracted images. The astrometric errors are estimated by taking the dispersions of the centroids of the transients in the subtracted images.

The main science objective of the HSC-SSP transient survey is Type~Ia SN cosmology at $1\lesssim z \lesssim 1.5$, but we conduct SHIZUCA with the same data to search for SNe at $z\gtrsim 1.5$.  The high-redshift SNe and SN candidates reported in this paper are selected based primarily on the COSMOS2015 photometric redshifts \citep{laigle2016cosmos2015} as well as the photometric redshifts estimated by the \texttt{MIZUKI} code \citep{tanaka2015mizuki} that is introduced below. Some high-redshift SN candidates have follow up spectroscopy using LRIS on the Keck-I telescope at W.M. Keck Observatory \citep{oke95lris}. The Keck/LRIS spectra are reported in the accompanying paper \citepalias{curtin2017hscspec}.  Table~\ref{tab:listofsne} presents the list of the transients presented in this paper with their host galaxy properties.

\begin{figure}
 \begin{center}
  \includegraphics[width=\columnwidth]{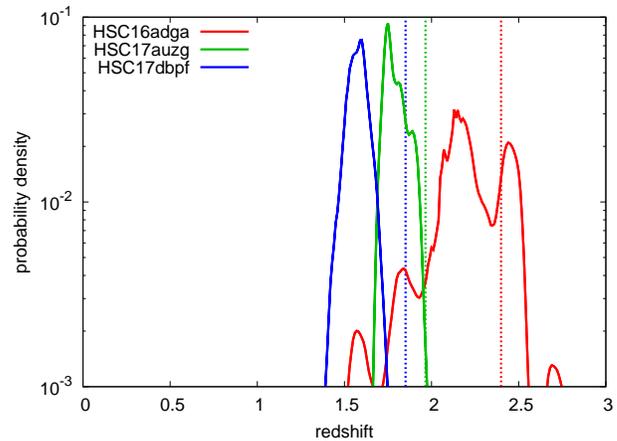}  
 \end{center}
\caption{
PDF of the photometric redshifts of the host galaxies of the spectroscopically-confirmed high-redshift SNe estimated by \texttt{MIZUKI} using the HSC and COSMOS2015 photometry of the host galaxies. The confirmed redshifts are shown with the vertical dotted lines. The COSMOS2015 photo-$z$ are $2.26^{+0.25}_{-0.30}$ (HSC16adga), $1.65^{+0.06}_{-0.08}$ (HSC17auzg), and $2.25^{+0.08}_{-0.53}$ (HSC17dbpf).
}\label{fig:speczphotozpdf}
\end{figure}

\begin{figure*}
 \begin{center}
  \includegraphics[width=1.6\columnwidth]{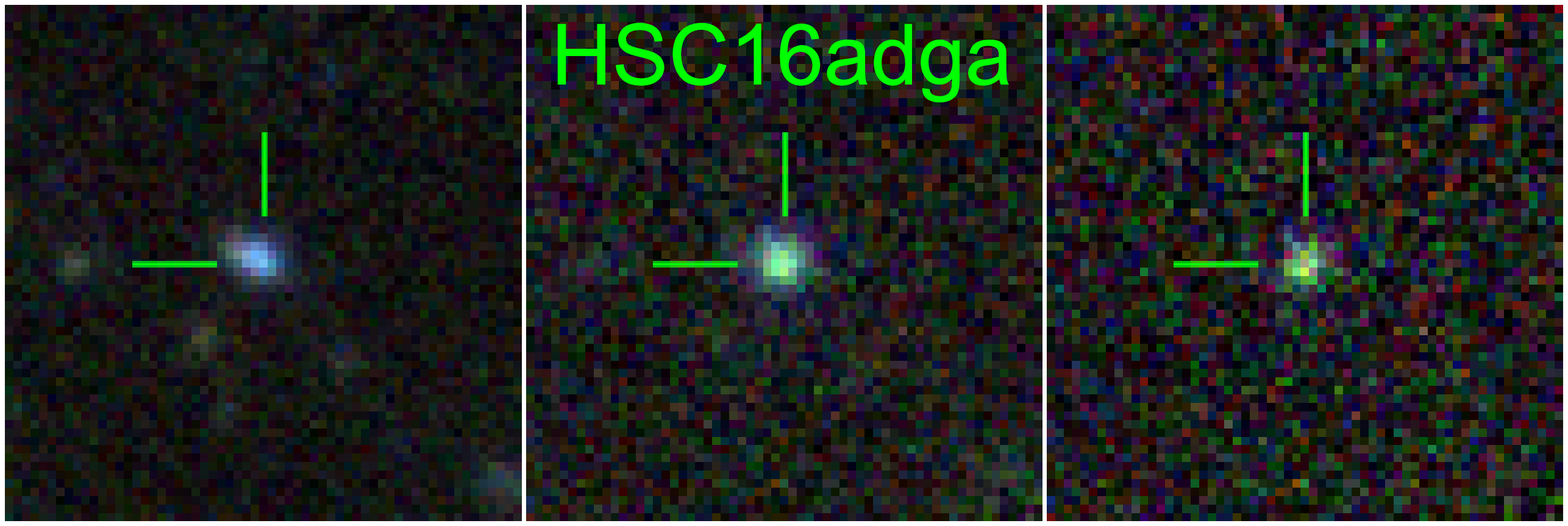}  \\ 
  \includegraphics[width=1.6\columnwidth]{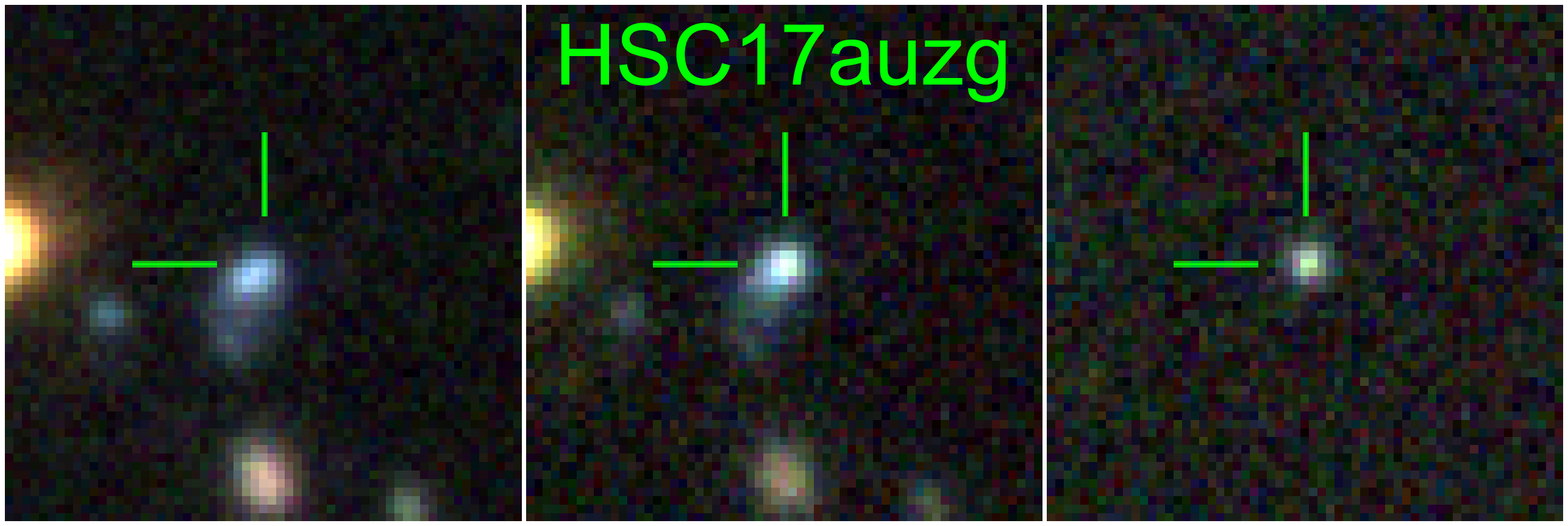}  \\
  \includegraphics[width=1.6\columnwidth]{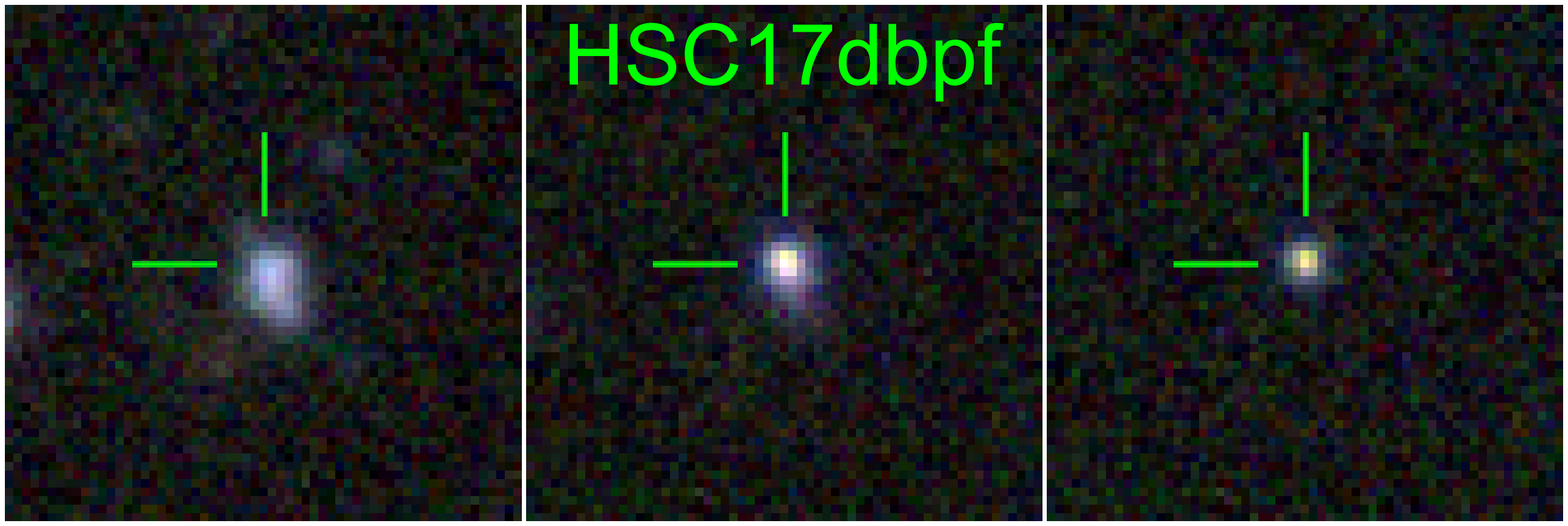}  
 \end{center}
\caption{
The images (10" x 10") of spectroscopically-confirmed high-redshift SNe and their host galaxies. The left panels show the reference images the middle panels show the images after the SN discovery at around the maximum brightness, and the right panels show their subtractions. Three filters ($i$, $r$, and $z$) are used to make the color composite. The SN locations are at the center of the images. North is up and east is left.
}\label{fig:speczfaces}
\end{figure*}

\section{High-redshift supernovae at around the redshift of 2}\label{sec:specconfhighz}
In this section, we focus on three SNe at $z\simeq 2$, i.e., HSC16adga ($z=2.399$), 
HSC17auzg ($z=1.965$), and HSC17dbpf ($z=1.851$). SN host galaxy physical properties, such as stellar mass and star formation rate (SFR) presented in this paper are estimated with the \texttt{MIZUKI} code \citep{tanaka2015mizuki}.
\texttt{MIZUKI} estimates the photometric redshifts of galaxies by using the Bayesian priors on their physical properties. It can estimate the photometric redshifts and the galaxy properties (stellar mass and SFR) simultaneously in a self-consistent way. \texttt{MIZUKI} is shown to provide good estimates for the host galaxy properties with smaller dispersions than the previous methods, for example. We refer \citet{tanaka2015mizuki} for more detailed descriptions of the code and its performance.
The photometric redshifts and the host galaxy properties are obtained by using all the available photometry in COSMOS2015 and HSC. The probability distribution functions (PDFs) of the estimated photometric redshifts for the host galaxies of the three SNe at $z\simeq 2$ are presented in Fig.~\ref{fig:speczphotozpdf}. The estimated galaxy properties from \texttt{MIZUKI} in this section are obtained by assuming the spectroscopically confirmed redshifts. The image cutouts of the SNe are shown in Fig.~\ref{fig:speczfaces}. All the SN photometric data are summarized in the Appendix.

\begin{figure}
 \begin{center}
  \includegraphics[width=\columnwidth]{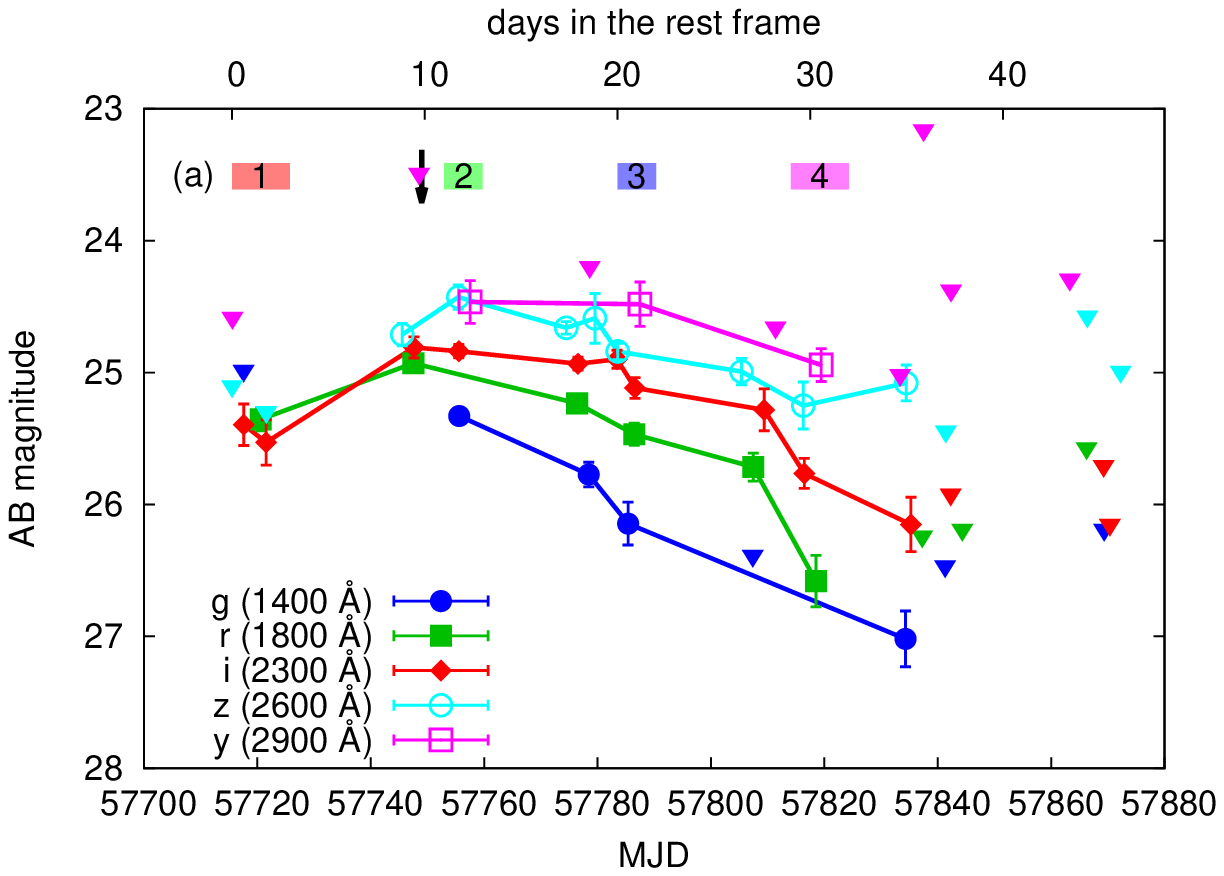}  
  \includegraphics[width=\columnwidth]{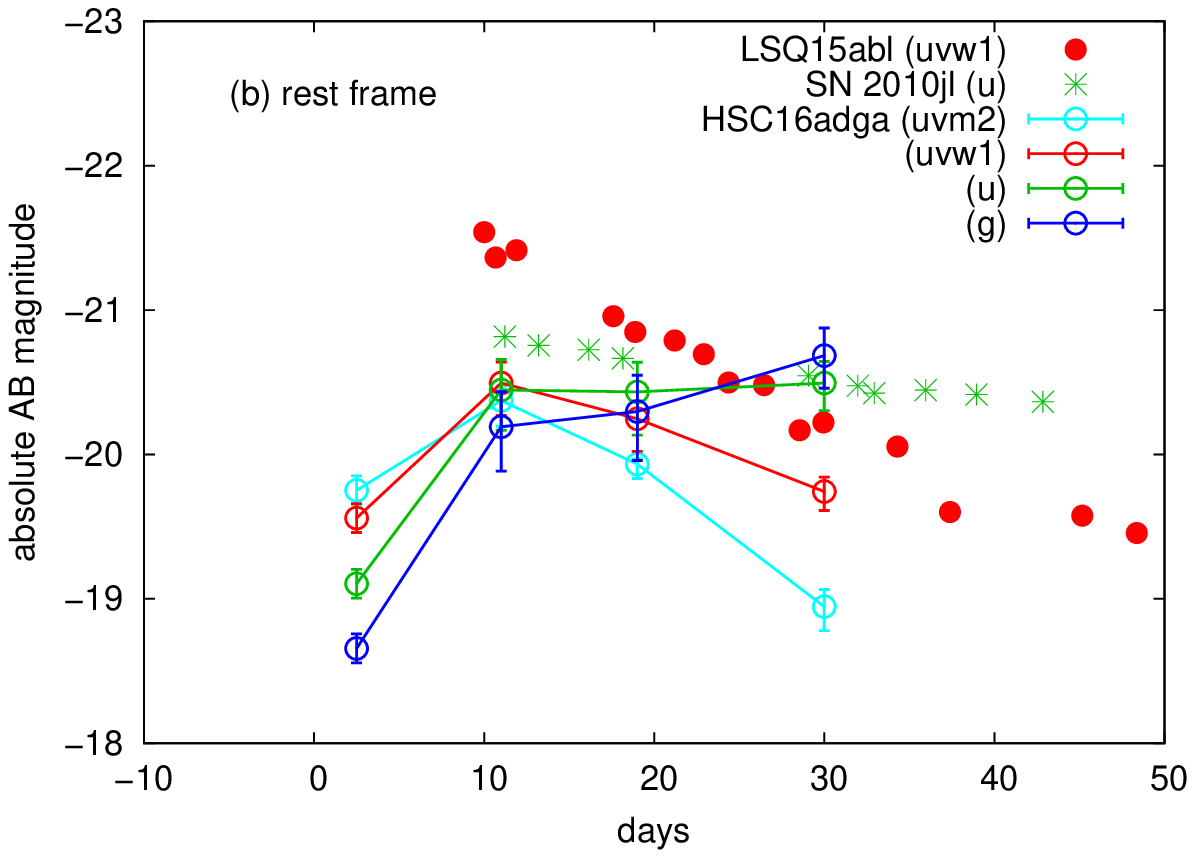}
 \end{center}
\caption{
LCs of HSC16adga. Detections with the significance of more than $5\sigma$ are presented with the $1\sigma$ error and triangles show the $5\sigma$ detection limits. Detections are connected with lines. (a) Observed LCs (Table~\ref{tab:hsc16adgaphotometory}).
The top x axis shows the time after the discovery at $z=2.399$. The central wavelength at $z=2.399$ for each filter is shown.  The regions with numbers on top indicate the epochs when the SEDs are shown in Fig.~\ref{fig:hsc16adgaSED} with which the rest-frame LCs in the panel (b) are obtained. The arrow shows when the spectrum is taken by Keck/LRIS. (b) The rest-frame ultraviolet and optical LCs of HSC16adga. The $K$ correction is based on the blackbody SED fits presented in Fig.~\ref{fig:hsc16adgaSED}. The errors include the blackbody fitting and photometric uncertainties. The LCs of Type~IIn SLSN LSQ15abl \citep{brown2014swiftsncat} and the luminous Type~IIn SN 2010jl are shown for comparison \citep{fransson2014sn2010jl}. The time of the comparison LCs is shifted to match the LC peak of HSC16adga. 
}\label{fig:hsc16adgaLC}
\end{figure}

\begin{figure}
 \begin{center}
  \includegraphics[width=\columnwidth]{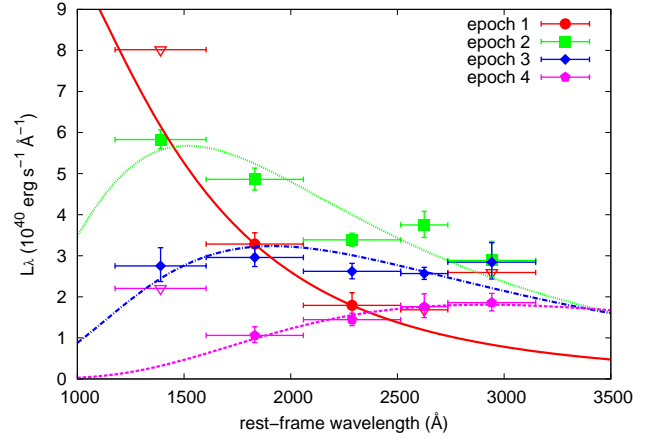}  
 \end{center}
\caption{
Rest-frame SEDs of HSC16adga estimated by the broad band photometry in the selected epochs shown in Fig.~\ref{fig:hsc16adgaLC}a. Open triangles indicate upper limits. The SEDs from blackbody temperatures ($T_\mathrm{BB}$) and radii ($R_\mathrm{BB}$)
that provide the best fit to the SEDs are shown 
(red solid line for the epoch 1: $T_\mathrm{BB}=36,000~\mathrm{K}$ and $R_\mathrm{BB}=3.3\times 10^{14}~\mathrm{cm}$ [no degree-of-freedom (dof)], green dotted line: $T_\mathrm{BB}=19,000~\mathrm{K}$ and $R_\mathrm{BB}= 1.2 \times 10^{15}~\mathrm{cm}$ [$\chi^2/\mathrm{dof}=1.15/3$], blue dot-dashed line: $T_\mathrm{BB}=15,000~\mathrm{K}$ and $R_\mathrm{BB}=1.6\times 10^{15}~\mathrm{cm}$ [$\chi^2/\mathrm{dof}=0.72/3$], and pink dashed line: $T_\mathrm{BB}=10,000~\mathrm{K}$ and $R_\mathrm{BB}=3.4\times 10^{15}~\mathrm{cm}$ [$\chi^2/\mathrm{dof}=0.02/2$]).
}\label{fig:hsc16adgaSED}
\end{figure}

\subsection{HSC16adga (SN~2016jhm)}\label{sec:hsc16adga}
HSC16adga (SN~2016jhm) was discovered shortly after the beginning of the survey at (RA, Dec) = (10:02:20.12, +02:48:43.3). It appeared in a galaxy with the COSMOS photo-$z$ of $2.26^{+0.25}_{-0.30}$ and the \texttt{MIZUKI} photo-$z$ peaking at $2.19$ (Fig.~\ref{fig:speczphotozpdf}). The \texttt{MIZUKI} photo-$z$ PDF has the second major peak at $z\simeq 2.45$. The SN spectroscopic redshift is $z=2.399\pm 0.004$ \citepalias{curtin2017hscspec} which is consistent with the broad photometric redshift probability distribution, especially with the second peak in the PDF in Fig.~\ref{fig:speczphotozpdf}. The SN location is $0.36"\pm 0.10"$ (2.9~kpc at $z=2.399$) away from the host galaxy center.
We estimate the host galaxy properties by fixing the redshift ($z=2.399$) using \texttt{MIZUKI}. It is estimated to be a star-forming galaxy with the stellar mass of $9.3^{+0.9}_{-1.3}\times 10^{9}~\Msun$ and the SFR of $ 13.1^{+0.3}_{-2.7}~\Msun~\mathrm{yr^{-1}}$.

Fig.~\ref{fig:hsc16adgaLC}a presents the observed light curves (LCs) of HSC16adga. The original data are summarized in Table~\ref{tab:hsc16adgaphotometory}.  Fig.~\ref{fig:hsc16adgaSED} is the rest-frame spectral energy distribution (SED) obtained with the broad band photometry at the selected epochs shown in Fig.~\ref{fig:hsc16adgaLC}a. We do not take any host galaxy extinction into account in this paper.
Since at early epochs the photometric temperatures are higher than $\simeq 20,000~\mathrm{K}$, the peak of the SEDs is not constrained by our optical photometry.
Afterward, the photospheric temperatures gradually cool to $\simeq 15,000~\mathrm{K}$ (20~days) and $\simeq 10,000~\mathrm{K}$ (30~days). The temperature evolution is consistent with that of luminous Type~IIn SNe \citep[e.g.,][]{fassia2000sn1998sphoto}.

Using the results of the blackbody SED fitted to the broad band photometry, we estimate the rest-frame ultraviolet (the $uvm2$ and $uvw1$ filters from \textit{Swift}) and optical (the $u$ band filter from the Sloan Digital Sky Survey and the $g$ band from HSC) LCs as shown in Fig.~\ref{fig:hsc16adgaLC}b.
The $u$ band magnitude of HSC16adga is similar to that of SN~2010jl, although the LC of HSC16adga is probably still rising. The $g$ band magnitude is getting more luminous and it is about to reach $-21$~mag. Although HSC16adga does not reach exactly $-21$~mag in optical, which was used to define SLSNe under one criterion \citep{gal-yam2012slsnrev}, it is currently known that many SLSNe, which are spectroscopically classified nowadays, do not reach $-21$~mag in optical \citep[e.g.,][]{decia2018ptfslsn,quimby2018slsn}. These facts, as well as the proximity to $-21$~mag, lead us to claim that HSC16adga is likely a SLSN.

\begin{figure}
 \begin{center}
  \includegraphics[width=\columnwidth]{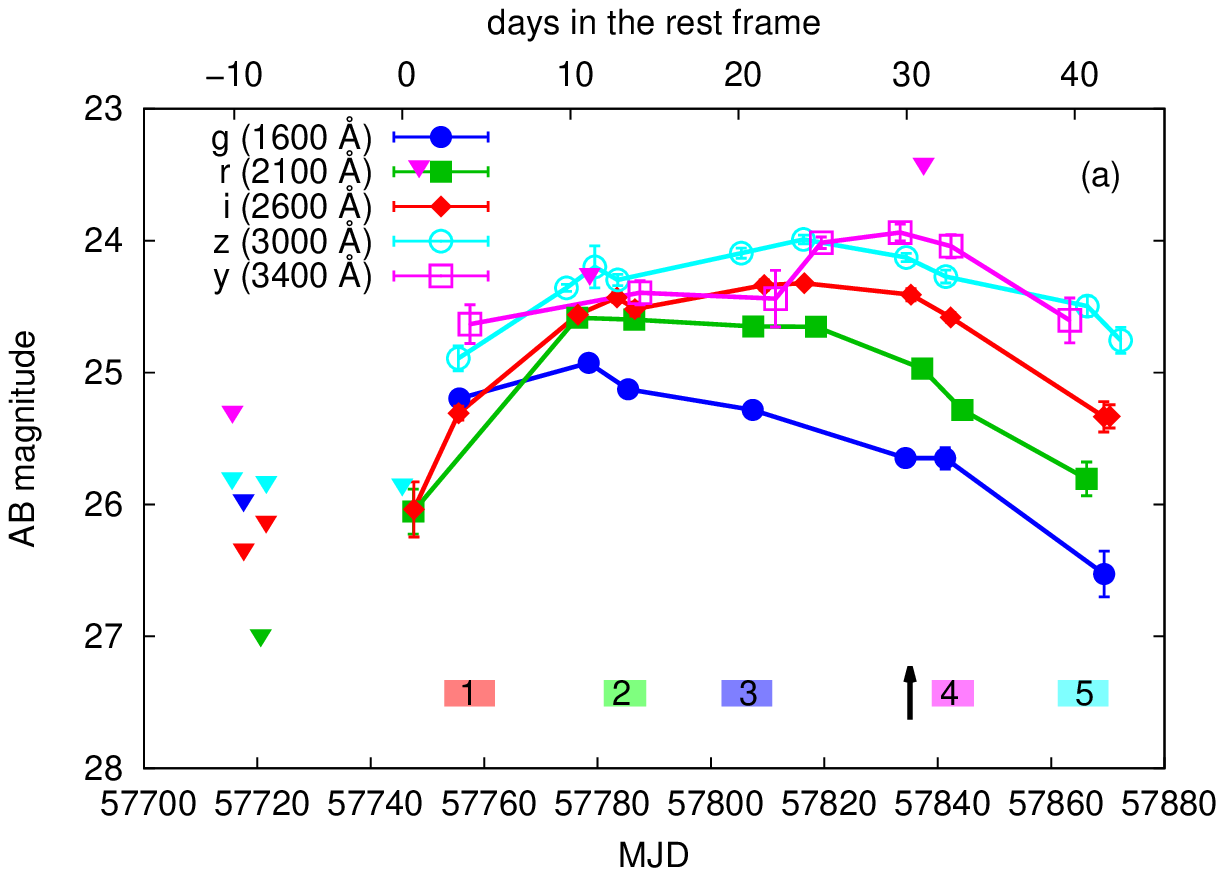}  
  \includegraphics[width=\columnwidth]{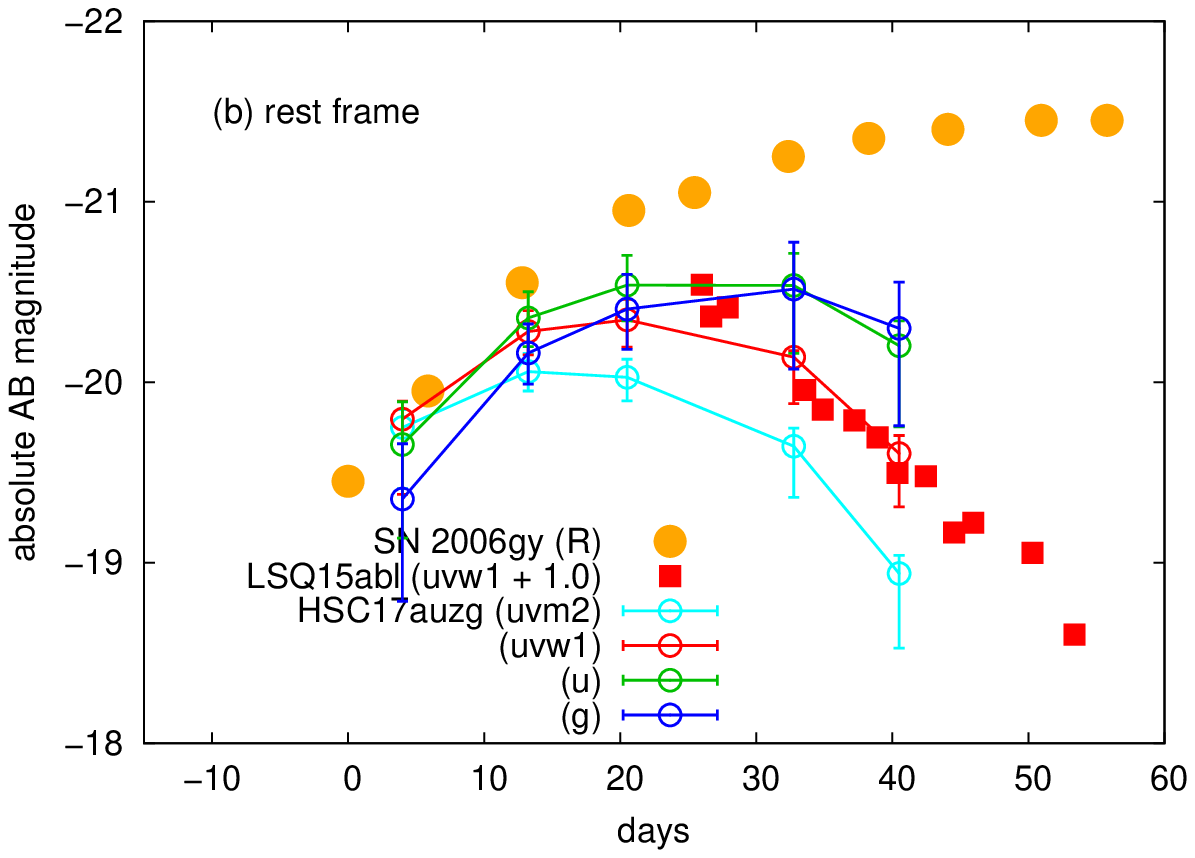}
 \end{center}
\caption{
LCs of HSC17auzg. Detections with the significance of more than $5\sigma$ are presented with the $1\sigma$ error and triangles show the $5\sigma$ detection limits. Detections are connected with lines. (a) Observed LCs (Table~\ref{tab:hsc17auzgphotometory}).
The top x axis shows the time after the discovery at $z=1.965$. The central wavelength at $z=1.965$ for each filter is shown. The regions with numbers on top indicate the epochs when the SEDs are shown in Fig.~\ref{fig:hsc17auzgSED} with which the rest-frame LCs in the panel (b) are obtained. The arrow shows when the spectrum is taken by Keck/LRIS. (b) The rest-frame ultraviolet and optical LCs of HSC17auzg. The $K$ correction is based the blackbody SED fits presented in Fig.~\ref{fig:hsc17auzgSED}. The errors include the blackbody fitting and photometric uncertainties. The LCs of Type~IIn SLSN 2006gy \citep{smith2007sn2006gyearly} and Type~IIn SLSN LSQ15abl \citep{brown2014swiftsncat} are shown for comparison. The time of the comparison LCs is shifted to match HSC17auzg. The magnitude of LSQ15abl is shifted by 1~mag so that the LC shape can be easily compared with that of HSC17auzg.
}\label{fig:hsc17auzgLC}
\end{figure}

\begin{figure}
 \begin{center}
  \includegraphics[width=\columnwidth]{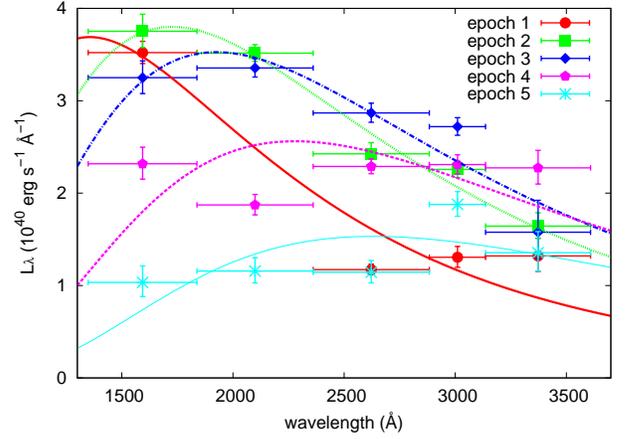}  
 \end{center}
\caption{
Rest-frame SEDs of HSC17auzg estimated by the broad band photometry in the selected epochs shown in Fig.~\ref{fig:hsc17auzgLC}a. The SEDs from several blackbody temperatures and radii
that provide the best fit to the SEDs are shown
 (red thick solid line: $T_\mathrm{BB}=21,000~\mathrm{K}$ and $R_\mathrm{BB}=7.2\times 10^{14}~\mathrm{cm}$ [$\chi^2/\mathrm{dof}=0.42/2$], green dotted line: $T_\mathrm{BB}=17,000~\mathrm{K}$ and $R_\mathrm{BB}=1.3\times 10^{15}~\mathrm{cm}$ [$\chi^2/\mathrm{dof}=0.09/3$], blue dot-dashed line: $T_\mathrm{BB}=15,000~\mathrm{K}$ and $R_\mathrm{BB}=1.7\times 10^{15}~\mathrm{cm}$ [$\chi^2/\mathrm{dof}=0.26/3$], pink dashed line: $T_\mathrm{BB}=13,000~\mathrm{K}$ and $R_\mathrm{BB}=2.2\times 10^{15}~\mathrm{cm}$ [$\chi^2/\mathrm{dof}=0.91/3$], and cyan thin solid line: $T_\mathrm{BB}=11,000~\mathrm{K}$ and $R_\mathrm{BB}=2.4\times 10^{15}~\mathrm{cm}$ [$\chi^2/\mathrm{dof}=0.44/3$]).
}\label{fig:hsc17auzgSED}
\end{figure}

\subsection{HSC17auzg (SN~2016jhn)}\label{sec:hsc17auzg}
HSC17auzg (SN~2016jhn) was first detected on 23 Dec 2016 at (RA, Dec) = (09:59:00.42, +02:14:20.8) in the $z$ band (Fig.~\ref{fig:speczfaces}). It appeared in a galaxy with the COSMOS photo-$z$ of $1.65^{+0.06}_{-0.08}$ and the \texttt{MIZUKI} photo-$z$ centering at $1.78$ (Fig.~\ref{fig:speczphotozpdf}). The spectroscopic follow-up observations confirmed the redshift of $1.965\pm 0.004$ \citepalias{curtin2017hscspec}.
The spectroscopic redshift is at the upper limit allowed by the photometric redshift PDF.
The SN is at $0.78"\pm 0.05"$ (6.5~kpc at $z=1.965$) away from the host galaxy center.
The host galaxy is estimated to have the stellar mass of $3.0^{+0.1}_{-0.3}\times 10^{10}~\Msun$ and the SFR of $33.8^{+8.1}_{-0.8}~\Msun~\mathrm{yr^{-1}}$ by \texttt{MIZUKI} using the spectroscopically confirmed redshift.

The LCs of HSC17auzg are shown in Fig.~\ref{fig:hsc17auzgLC} (see Table~\ref{tab:hsc17auzgphotometory} for the data). After the first detection, its brightness continued to increase for about 3~months in the $z$~band, while the rise times in the bluer bands are shorter (Fig.~\ref{fig:hsc17auzgLC}a). The SED evolution obtained by the broad band photometry is presented in Fig.~\ref{fig:hsc17auzgSED}. The blackbody temperatures evolve from above $\simeq 21,000~\mathrm{K}$ to $\simeq 11,000~\mathrm{K}$ in about 40~days. The corresponding blackbody radii are $\simeq 7\times 10^{14}~\mathrm{cm}$ to $\simeq 2\times 10^{15}~\mathrm{cm}$. 

Fig.~\ref{fig:hsc17auzgLC}b shows the rest-frame ultraviolet and optical LCs of HSC17auzg which are obtained by using the blackbody SEDs in Fig.~\ref{fig:hsc17auzgSED}. The LCs are compared with those of SLSN~IIn 2006gy \citep[e.g.,][]{smith2007sn2006gyearly} and Type~IIn SLSN LSQ15abl \citep{brown2014swiftsncat}. Although the rest-frame ultraviolet LCs of SN~2006gy do not exist, its optical LC rising rate is found to be consistent with HSC17auzg. The LC declining rate of HSC17auzg is consistent with that of LSQ15abl. The peak absolute magnitudes in the wavelength range presented in the figure do not reach $-21~\mathrm{mag}$, which classically separates SLSNe from the other SNe by one definition \citep{gal-yam2012slsnrev}. However, it is currently known that SLSNe do not necessarily reach $-21~\mathrm{mag}$ \citep[e.g.,][]{decia2018ptfslsn,quimby2018slsn}. The LCs in the bluer bands are found to start declining earlier and the LCs in the redder optical bands may actually keep rising in HSC17auzg (Fig.~\ref{fig:hsc17auzgLC}b). Therefore, we believe that HSC17auzg is likely a SLSN.

\begin{figure}
 \begin{center}
  \includegraphics[width=\columnwidth]{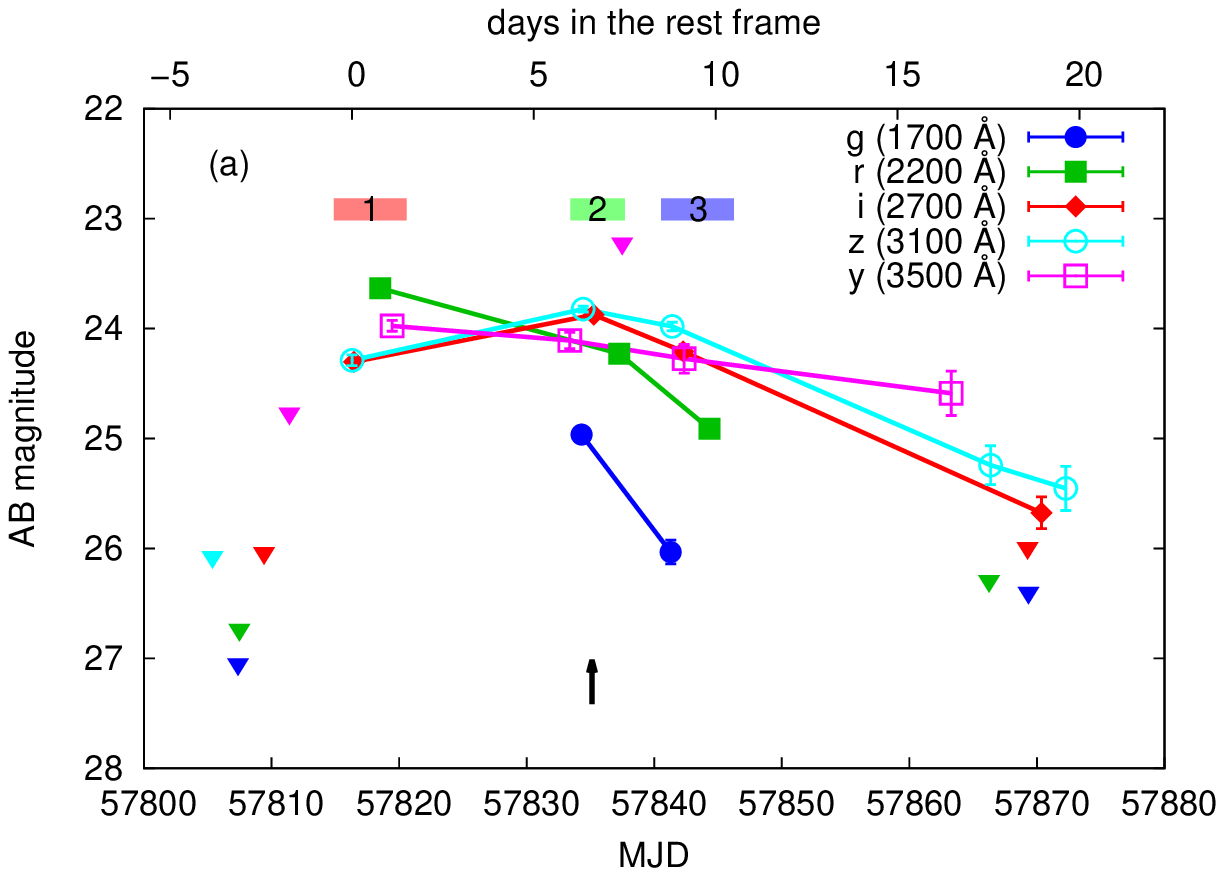}  
  \includegraphics[width=\columnwidth]{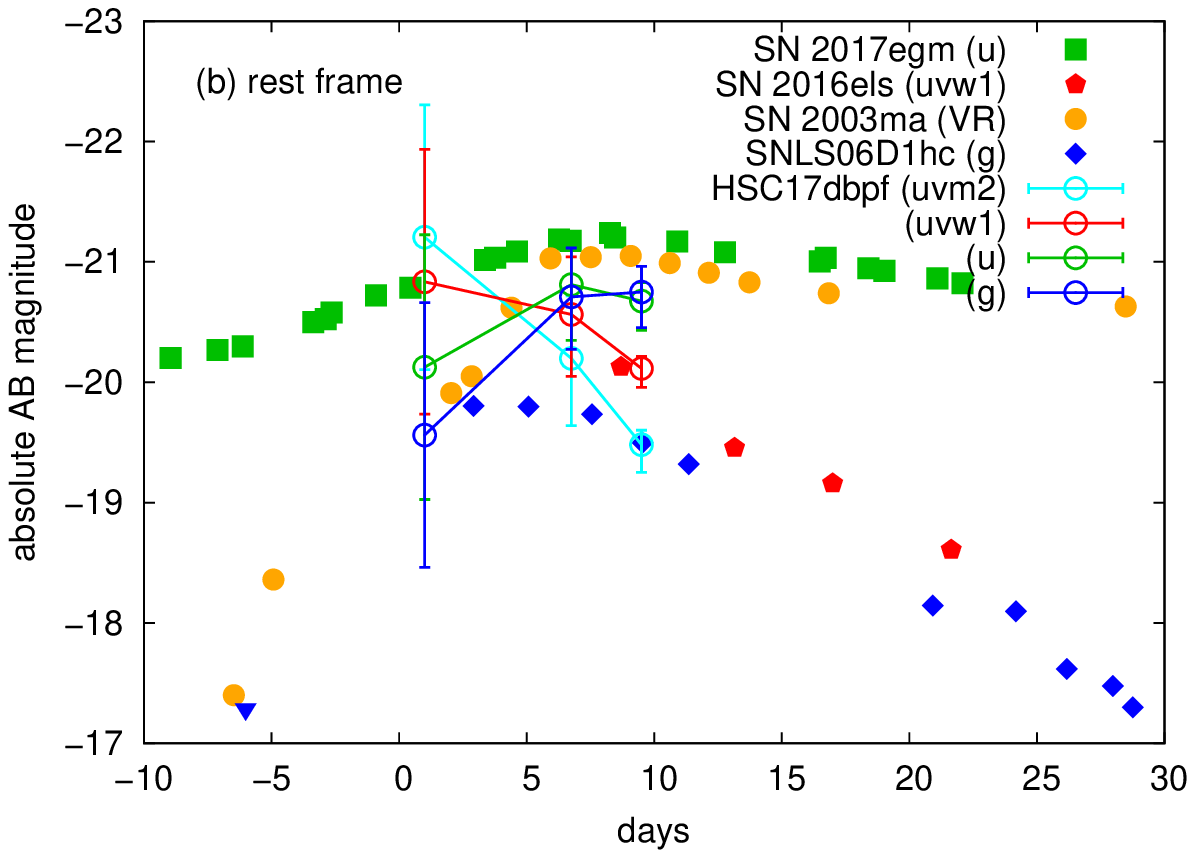}
 \end{center}
\caption{
LCs of HSC17dbpf. Detections with the significance of more than $5\sigma$ are presented with the $1\sigma$ error and triangles show the $5\sigma$ detection limit. Detections are connected with lines. (a) Observed LCs (Table~\ref{tab:hsc17dbpfphotometory}).
The top x axis shows the time after the discovery at $z=1.851$. The central wavelength at $z=1.851$ for each filter is shown. The regions with numbers on top indicate the epochs when the SEDs are shown in Fig.~\ref{fig:hsc17dbpfSED} with which the rest-frame LCs in the panel (b) are obtained. The arrow shows when the spectrum is taken by Keck/LRIS. (b) The rest-frame ultraviolet and optical LCs of HSC17dbpf. The $K$ correction is based on the blackbody SED fits presented in Fig.~\ref{fig:hsc17dbpfSED}. The errors include the blackbody fitting and photometric uncertainties. The LCs of Type~I SLSNe 2017egm \citep{bose2017sn2017egm} and 2016els \citep{brown2014swiftsncat}, the rapidly-rising Type~IIn SLSN~2003ma \citep{rest2011sn2003ma}, and the luminous SN SNLS06D1hc \citep{arcavi2016slsnsngap} are shown for comparison. The time of the comparison LCs is shifted to match HSC17bdpf.
}\label{fig:hsc17dbpfLC}
\end{figure}

\begin{figure}
 \begin{center}
  \includegraphics[width=\columnwidth]{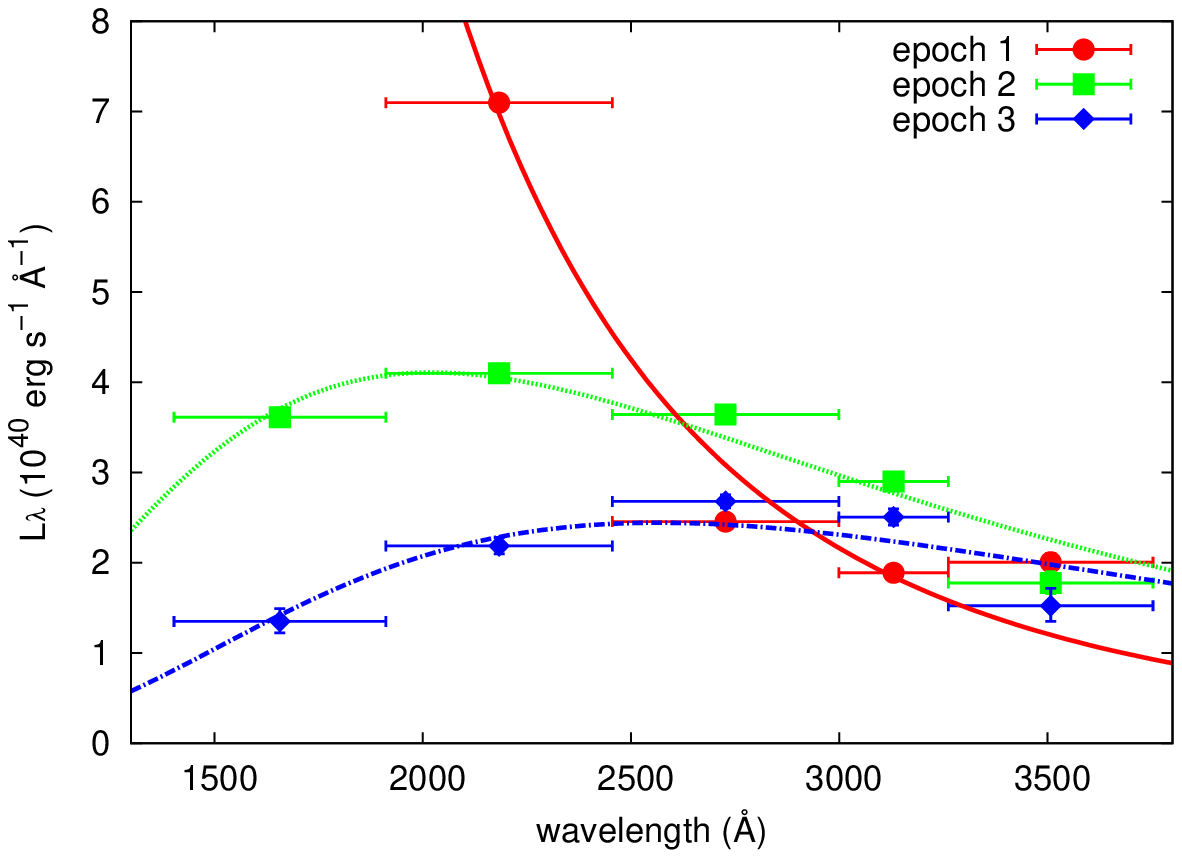}  
 \end{center}
\caption{
Rest-frame SEDs of HSC17dbpf estimated by the broad band photometry in the selected epochs shown in Fig.~\ref{fig:hsc17dbpfLC}a. The SEDs from several blackbody temperatures and radii 
that provide the best fit to the SEDs are shown
(red solid line: $T_\mathrm{BB}=98,000~\mathrm{K}$ and $R_\mathrm{BB}=2.6\times 10^{14}~\mathrm{cm}$ [$\chi^2/\mathrm{dof}=1.06/2$], green dotted line: $T_\mathrm{BB}=14,000~\mathrm{K}$ and $R_\mathrm{BB}=2.0\times 10^{15}~\mathrm{cm}$ [$\chi^2/\mathrm{dof}=0.33/3$], and blue dot-dashed line: $T_\mathrm{BB}=11,000~\mathrm{K}$ and $R_\mathrm{BB}=2.9\times 10^{15}~\mathrm{cm}$ [$\chi^2/\mathrm{dof}=0.36/3$]).
}\label{fig:hsc17dbpfSED}
\end{figure}

\begin{figure*}
 \begin{center}
  \includegraphics[width=1.6\columnwidth]{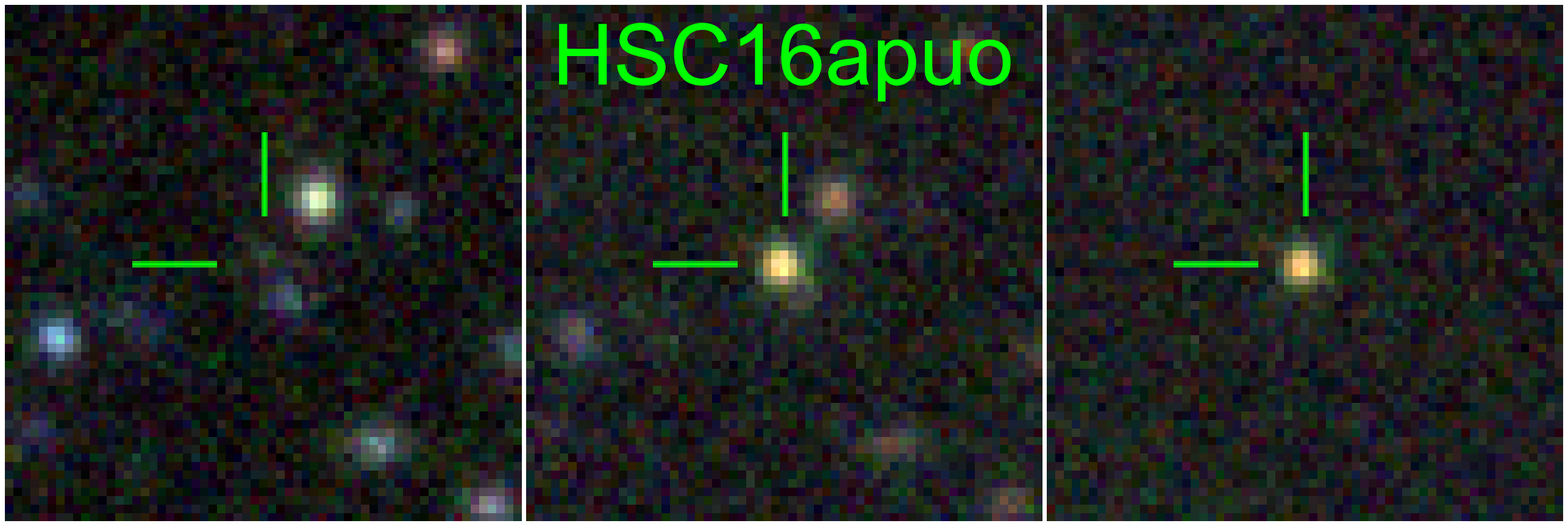}  \\
  \includegraphics[width=1.6\columnwidth]{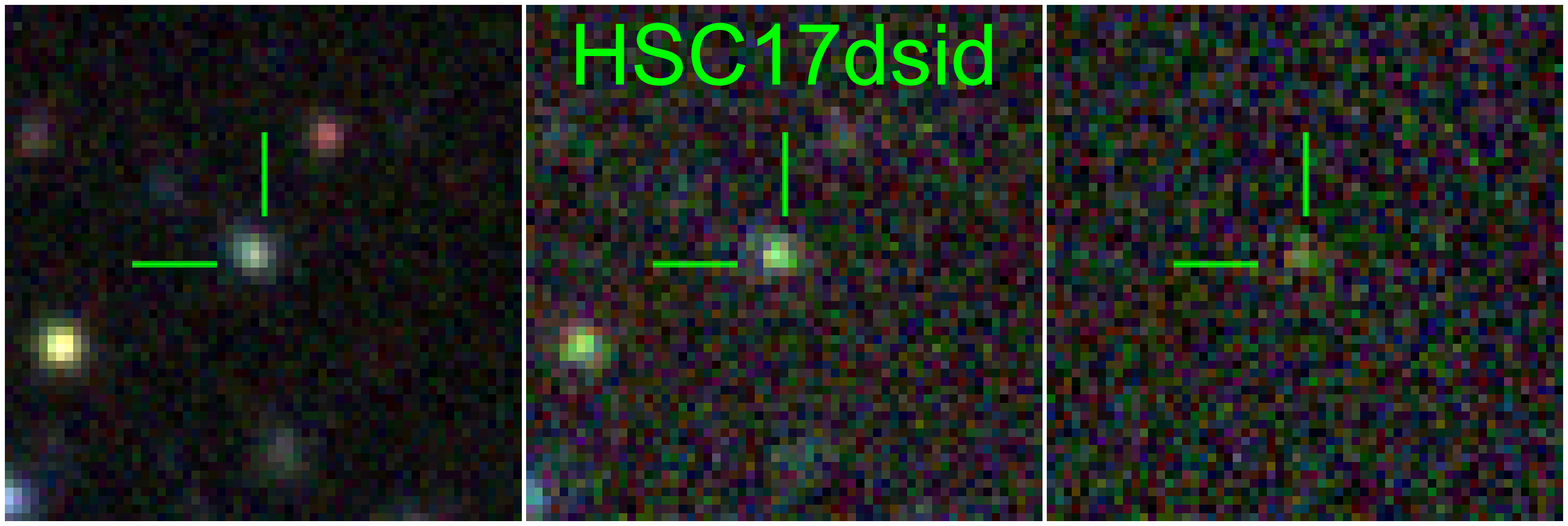}  
 \end{center}
\caption{
The images (10" x 10") of high-redshift SN candidates. The left panels show the reference images, the middle panels show images after the SN discovery, and the right panels show their subtractions. Three filters ($i$, $r$, and $z$) are used to make the color composite. The SN locations are at the center of the images. For HSC17dsid, the image cutouts for all the epochs are available in the Appendix. North is up and east is left.
}\label{fig:photozfaces}
\end{figure*}

\begin{figure}
 \begin{center}
  \includegraphics[width=\columnwidth]{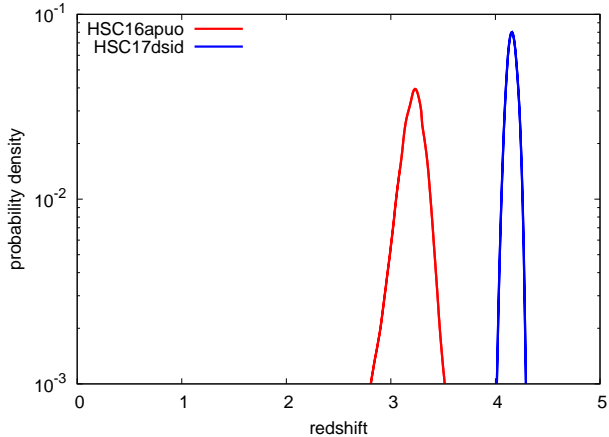}  
 \end{center}
\caption{
Photometric redshifts of the host galaxies of the high-redshift SN candidates estimated by \texttt{MIZUKI} using the HSC and COSMOS2015 photometry of the host galaxies (Fig.~\ref{fig:hostsed}). The COSMOS2015 photo-$z$ are $2.82^{+0.47}_{-0.70}$ (HSC16apuo) and $4.20^{+0.09}_{-0.13}$ (HSC17dsid).
}\label{fig:photozphotozpdf}
\end{figure}

\begin{figure}
 \begin{center}
  \includegraphics[width=0.89\columnwidth]{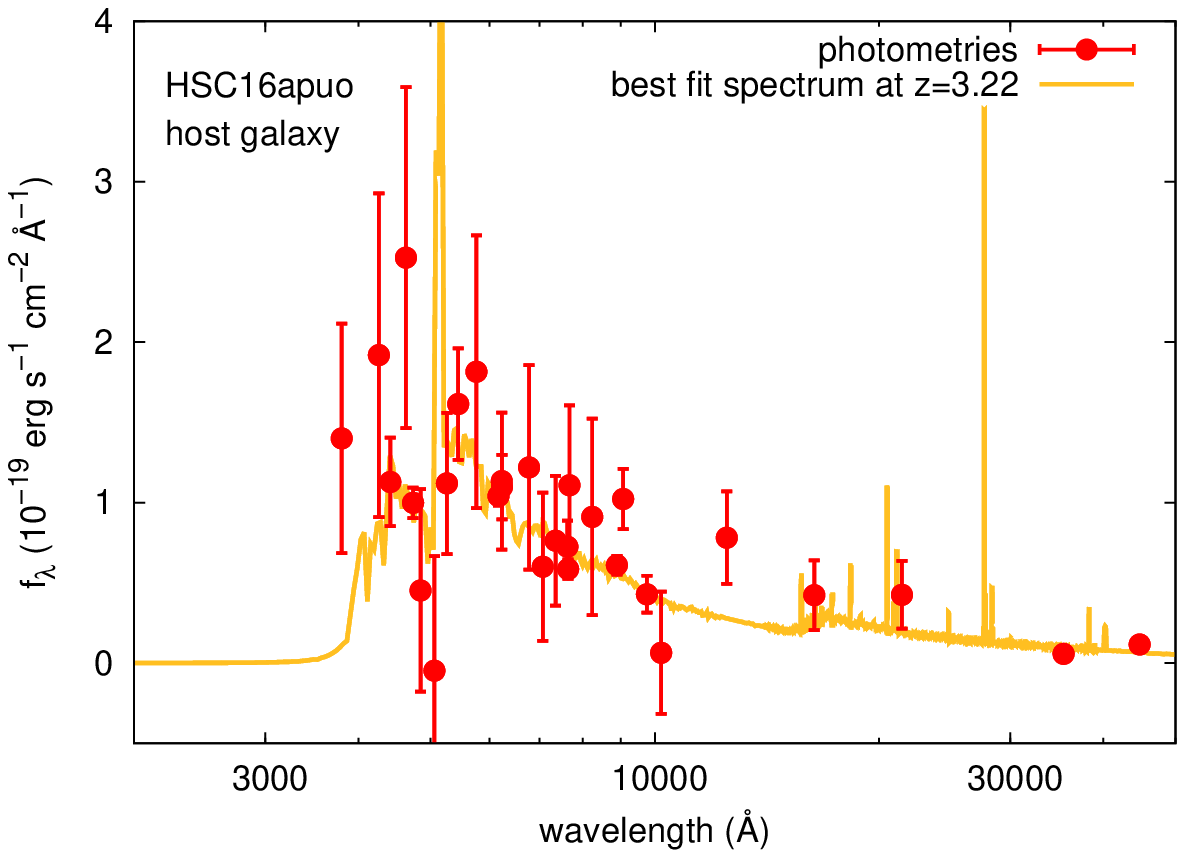}  
  \includegraphics[width=0.89\columnwidth]{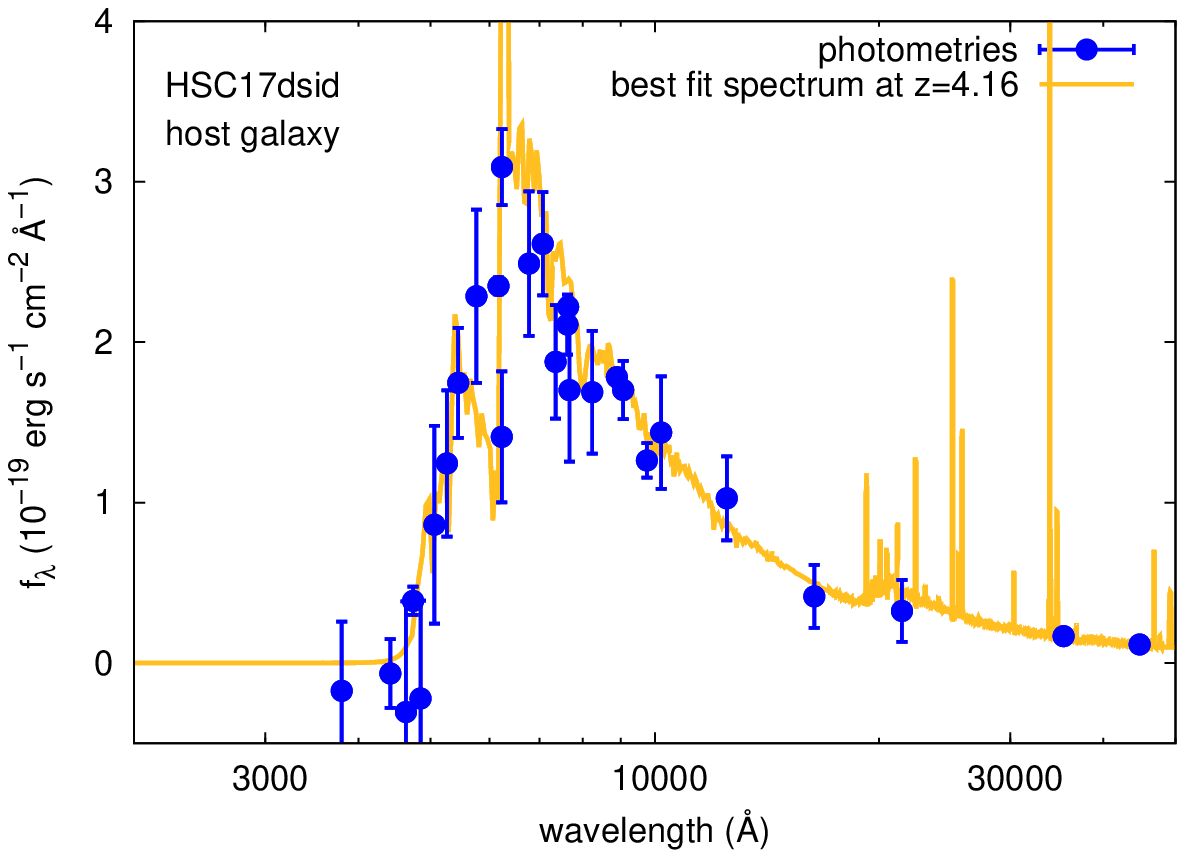}  
 \end{center}
\caption{
SEDs of the host galaxies of our high-redshift SN candidates with the best fit synthesized galaxy spectra obtained by \texttt{MIZUKI}. The photometric data are from HSC and COSMOS2015.
}\label{fig:hostsed}
\end{figure}

\subsection{HSC17dbpf (SN~2017fei)}\label{sec:hsc17dbpf}
The third spectroscopically-confirmed high-redshift SN is HSC17dbpf (SN~2017fei) at (RA, Dec) = (09:58:33.42, +01:59:29.7). It appeared in a galaxy with the COSMOS photo-$z$ of $2.25^{+0.08}_{-0.53}$ and the \texttt{MIZUKI} photo-$z$ centering at $1.58$ (Fig.~\ref{fig:speczphotozpdf}). The redshift of HSC17dbpf is confirmed to be $1.851\pm 0.004$ with the spectroscopic follow-up observation \citepalias{curtin2017hscspec}. 
Our spectroscopic redshift is higher than that estimated by the \texttt{MIZUKI} code, but it is in the 1$\sigma$ error range of the photometric redshift estimated by COSMOS.
The SN location is $0.58"\pm  0.08"$ (4.9~kpc at $z=1.851$) away from the host galaxy center.
\texttt{MIZUKI} estimates the host galaxy stellar mass of $9.2^{+0.2}_{-3.1}\times 10^{9}~\Msun$ and the SFR of $40.0^{+5.1}_{-1.1}~\Msun~\mathrm{yr^{-1}}$.

The LCs of HSC17dbpf are reported in Fig.~\ref{fig:hsc17dbpfLC}. The original data are in Table~\ref{tab:hsc17dbpfphotometory}. HSC17dbpf has the LC that evolves more rapidly than the other two high-redshift SNe we found.
The rest-frame SED evolution of HSC17dbpf is presented in Fig.~\ref{fig:hsc17dbpfSED}. The blackbody temperature in the first epoch is not well constrained, but the steep rise in the SED indicates a high temperature above 30,000~K. The blackbody temperature evolves to $\simeq 14,000~\mathrm{K}$ on the second epoch and then goes down to $\simeq 11,000~\mathrm{K}$ on the third epoch. The early high temperature is consistent with that found in SN~2003ma \citep{rest2011sn2003ma} and Type~I SLSNe \citep[e.g.,][]{nicholl2016bumpubiq}. However, Type~I SLSN spectra below around 3000~\AA\ often deviate from a blackbody \citep[e.g.,][]{yan2017gaia16apduv,yan2017sn2017egmuvspec}.

Fig.~\ref{fig:hsc17dbpfLC}b presents the ultraviolet and optical LCs of HSC17dbpf in the rest frame. The rest-frame LCs are obtained by using the blackbody SEDs shown in Fig.~\ref{fig:hsc17dbpfSED}. The peak absolute magnitudes are consistent within the uncertainty to $-21$~mag. The ultraviolet magnitudes of HSC17dbpf are similar to those of some Type~I SLSNe and it is also likely that the LCs in the redder optical bands are still rising. Therefore, we claim that HSC17dbpf is a SLSN. The rapid LC rise is consistent with that of Type~IIn SLSN 2003ma, although our rest-frame absolute magnitude estimates have large errors at the beginning. Such a rapid rise is not found in optical LCs of Type~I SLSNe, but the rest-ultraviolet LCs of Type~I SLSNe to compare are lacking. It is also interesting to note that the rapid evolution is consistent with the rapidly evolving transients from the Dark Energy Survey recently reported by \citet{pursiainen2018rapiddes} as well as those found by our previous HSC survey \citep{tanaka2016hscrapidrise}. The recent discovery of the Type~I SLSN 2017egm at $z=0.031$ \citep[e.g.,][]{bose2017sn2017egm,nicholl2017sn2017egm} provided an opportunity to obtain the rest ultraviolet photometry of the Type~I SLSN but the rise is slower than that of HSC17dbpf \citep{bose2017sn2017egm}. The rapid rise is also found in some SNe in the luminosity range between SNe and SLSNe \citep{arcavi2016slsnsngap}, but HSC17dbpf is brighter (Fig.~\ref{fig:hsc17dbpfLC}b). The rapid decline of HSC17dbpf is consistent with that of Type~I SLSN 2016els, although the rise of this SN was not observed \citep{brown2014swiftsncat}.

\subsection{Other $z\sim 2$ SN candidates}
We have reported spectroscopically-confirmed $z\simeq 2$ SNe so far. As presented at the beginning of this section, the $z\sim 2$ candidates are selected based on the photometric redshift of the transients' host galaxies. We obtained about 10 SLSN candidates at $1.5\lesssim z\lesssim 2.5$ with this method, but we often did not have opportunities to conduct spectroscopic follow-up observations for them because of weather conditions and telescope availability.
The priorities for spectroscopic follow-up observations on a given night, which is not Target-of-Opportunity but classically scheduled in the case of the Keck follow up observations, are decided by the LC information, the apparent magnitudes of the candidates, and the uncertainties in the photometric redshift estimates. We first select SN candidates with at least a few epochs of LC information to see if they are consistent with being SNe. Then, we sort the candidates with host galaxy photometric redshifts. The higher redshift SN candidates with small redshift uncertainties had higher priority for the spectroscopic follow up. Then, we made the final decision based on the apparent magnitudes which should be brighter than $\sim 25.5~\mathrm{mag}$ for the Keck/LRIS follow up or $\sim 24.5~\mathrm{mag}$ for Gemini/GMOS-S follow up to take a spectrum in a few hours. For example, HSC16apuo had a highest priority to follow for a while but bad weather prevented us from taking its spectrum when it was bright enough.
The remaining candidates and their LCs will be reported elsewhere after the completion of the survey.

\section{High-redshift supernova candidates beyond the redshift of 3}\label{sec:highzcand}
We additionally report two high-redshift SN candidates, one at $z\sim 3$ and the other at $z\sim 4$, to demonstrate the capability of the SHIZUCA. Neither the SN spectra nor the host galaxy spectra are taken so far\footnote{While this paper had been finalized, we had an opportunity to take the host galaxy spectrum of HSC17dsid. It will be presented in a future paper.}. However, their host galaxy photometric redshifts suggest the high-redshift nature of the SN candidates. Fig.~\ref{fig:photozfaces} shows the images of these high-redshift SN candidates. For the $z\sim 4$ SN candidate (HSC17dsid), we show the image cutouts of all the observed epochs in Table~\ref{tab:hsc17dsidphotometory}. Fig.~\ref{fig:photozphotozpdf} presents the host galaxy photometric redshifts obtained by \texttt{MIZUKI}. Their SEDs and the best fit host galaxy synthetic spectra are shown in Fig.~\ref{fig:hostsed}.

\subsection{HSC16apuo (AT~2016jho)}\label{sec:hsc16apuo}
\begin{figure}
 \begin{center}
  \includegraphics[width=\columnwidth]{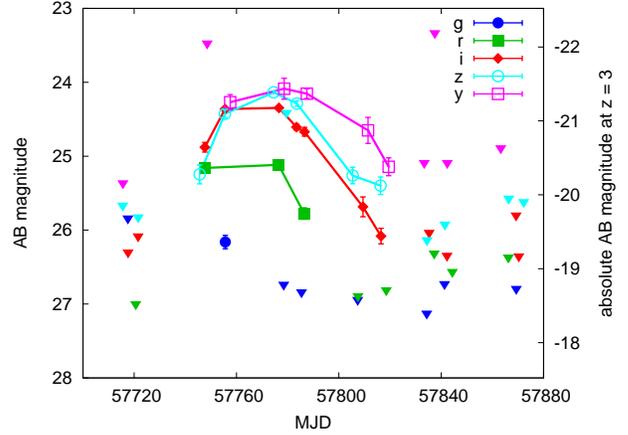}  
 \end{center}
\caption{
LCs of HSC16apuo (Table~\ref{tab:hsc16apuophotometory}). Detections with the significance of more than $5\sigma$ are presented with the $1\sigma$ error and triangles show the $5\sigma$ detection limit. Detections are connected with lines. The right axis shows the expected absolute magnitude at $z=3$ that includes the simple $K$ correction of $2.5\log (1+z)$ \citep{hogg2002kcorrections}.
}\label{fig:hsc16apuoLC}
\end{figure}

HSC16apuo (AT~2016jho) was discovered at (RA, Dec) = (10:01:29.42, +02:28:33.8) on 23 Dec 2016 in a galaxy with the COSMOS photo-$z$ of $2.82^{+0.47}_{-0.70}$ (Fig.~\ref{fig:photozfaces}).
It appeared $0.56"\pm 0.05"$ away from the host galaxy center, which is located in the south west of HSC16apuo. The corresponding physical distance at $z=3$ is 4.3~kpc. The PDF of the photometric redshift of the host galaxy estimated by \texttt{MIZUKI} is shown in Fig.~\ref{fig:photozphotozpdf}. The most probable redshift is 3.22 but the PDF has the relatively extended distribution ranging from $z\simeq 2.8$ to 3.5.
The SED of the host galaxy is presented in Fig.~\ref{fig:hostsed}. The relatively large photometric uncertainties lead to the uncertainty in the photometric redshift but its faintness and the small flux below $\sim 4000$~\AA\ support the photometric redshift of around 3 and reject the low redshift possibilities.
The estimated stellar mass and SFR of the host galaxy are $4.5^{+2.1}_{-1.3}\times 10^{9}~\Msun$ and the SFR of $6.0^{+2.5}_{-1.7}~\Msun~\mathrm{yr^{-1}}$, respectively. The best fit SED model presented in Fig.~\ref{fig:hostsed} has $\chi^2/\mathrm{dof} = 26.80/30$.
Assuming $z=3$, the SN is brighter than $-21$~mag at the peak (Fig.~\ref{fig:hsc16apuoLC}, see Table~\ref{tab:hsc16apuophotometory} for photometry data).

Although HSC16apuo appears to be at the edge of this galaxy, there is another galaxy at 1.9" towards the north west. The photometric redshift of this galaxy is degenerate and it could be either $z\sim 0.5$ or 3. 

\subsection{HSC17dsid (AT~2017fej)}\label{sec:hsc17dsid}
HSC17dsid (AT~2017fej) is the most distant SN candidate found in the survey so far. It was discovered at (RA, Dec) = (10:02:58.12, +02:13:04.1) on 26 Apr 2017 in a galaxy with the COSMOS photo-$z$ of $4.1974^{+0.09}_{-0.13}$ (Fig.~\ref{fig:photozfaces}). The SN location is $0.16"\pm 0.07"$ away (1.1~kpc at $z=4$) from the host galaxy center. The PDF of the photometric redshift of the host galaxy estimated by \texttt{MIZUKI} is shown in Fig.~\ref{fig:photozphotozpdf}. The most probable redshift is 4.16. The PDF has a narrow distribution and HSC17dsid is likely at $z\sim 4$. The host galaxy SED presented in Fig.~\ref{fig:hostsed} clearly shows the signature of the Lyman break at $\sim 5000$~\AA, which strongly supports the host redshift of $z\sim 4$.
The stellar mass and SFR of the host galaxy are estimated to be $3.4^{+4.8}_{-2.0}\times 10^{9}~\Msun$ and $24.2_{-24.2}^{+15.1}~\Msun~\mathrm{yr^{-1}}$, respectively. 
The best fit SED model presented in Fig.~\ref{fig:photozphotozpdf} has $\chi^2/\mathrm{dof} = 22.39/30$.

\begin{figure}
 \begin{center}
  \includegraphics[width=\columnwidth]{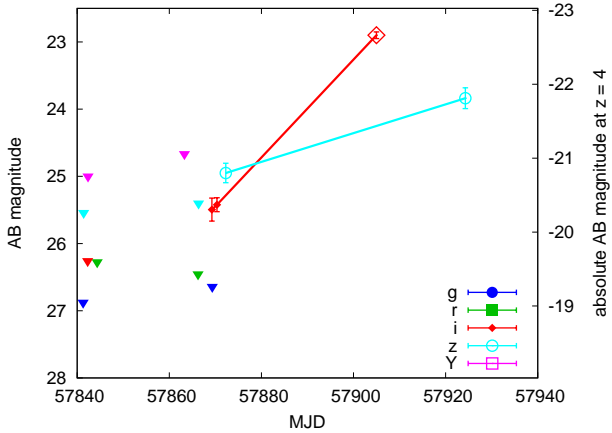}  
 \end{center}
\caption{
LCs of HSC17dsid (Table~\ref{tab:hsc17dsidphotometory}). Detections with the significance of more than $5\sigma$ are presented with the $1\sigma$ error and triangles show the $5\sigma$ detection limit. The $i$ band observation with Gemini/GMOS-S is shown with the red open diamond shape. Detections are connected with lines. The right axis shows the expected absolute magnitude at $z=4$ that includes the simple $K$ correction of $2.5\log (1+z)$ \citep{hogg2002kcorrections}.
}\label{fig:hsc17dsidLC}
\end{figure}

The LC of HSC17dsid is shown in Fig.~\ref{fig:hsc17dsidLC} (see also Table~\ref{tab:hsc17dsidphotometory}). HSC17dsid was discovered at the end of our transient survey period. However, additional photometry was obtained by Gemini/GMOS-S on 31 May 2017 (UT) with the $i$ band (these data are reduced following standard procedures; open diamond in Fig.~\ref{fig:hsc17dsidLC}). Because HSC17dsid was much brighter than the host galaxy (24.8~mag in the $i$ band) when it was observed with Gemini, we did not perform the image subtraction for the Gemini photometry and we use the aperture photometry. We also had an opportunity to take another $z$ band image with HSC on 20 June 2017 (UT). Assuming $z=4$, HSC17dsid reaches $-22.7~\mathrm{mag}$ at 1600~\AA\ in about 10~days. It is brighter than the ultraviolet-brightest SLSN currently known (Gaia16apd, \citealt{yan2017gaia16apduv,nicholl2017gaia16apd}), which could be due to a smaller metallicity and thus lower extinction from the ejecta.

\section{Discussion}\label{sec:discussion}
\subsection{Comparison with other high-redshift SLSNe}
Two SLSNe have been reported in the literature that are at the similar redshifts to our SLSNe, DES15E2mlf at $z=1.861$ \citep{pan2017slsnz1p861} and DES16C2nm at $z=1.998$ \citep{smith2017z2des}. Their peak optical magnitudes are at around 23~mag, which is brighter than the peak magnitudes of our SLSNe ($\sim 24$~mag). They are found in the shallower transient survey and HSC allowed us to find fainter SLSNe.

Our high-redshift SLSNe are fainter in the rest-frame ultraviolet wavelengths which we directly observe, as well as in the rest-frame optical bands whose luminosity is estimated in the previous section. While DES15E2mlf and DES16C2nm are brighter than $-22$~mag in optical, our SLSNe basically remain to be  fainter than $-22$~mag in the rest frame optical. The rest-frame optical peak magnitudes of SLSNe are usually between $-19$~mag and $-22$~mag \citep[e.g.,][]{nicholl2017sn2017egm,decia2018ptfslsn,lunnan2018pansslsn}. Therefore, our survey is capable of finding SLSNe with peak magnitudes fainter than $-21$~mag, while the previous surveys could only observe the most luminous end of the high-redshift SLSN populations.

\subsection{Approximate event rates}
We roughly estimate the event rates of SLSNe at high redshifts based on the three spectroscopically-confirmed SNe and SN candidates we present here.  Because the complete selection of high-redshift SN candidates from the survey has not yet been finished, our rate estimates are approximate. We have at least a few more good SN candidates at $z\sim 2$ whose redshifts are not spectroscopically confirmed. The actual number depends on the choice of criteria, which include detection sigma and number of filters exceeding the threshold, LC evolution, color evolution, interpolation over missing epochs and missing filter information, etc. In addition, our high-redshift SN candidates are selected based on the photometric redshifts of the host galaxies and we have not completed our investigation into how our selection methods affect our discoveries.  For example, we may have missed SLSNe without apparent host galaxies or with faint ones (see also Section~\ref{sec:hostgalaxies} for related discussion).  These remaining candidates and final rate estimates will be presented elsewhere.

The event rate can be expressed as 
$\sum_{i} (1+z_i)/\varepsilon_i VT,$
where $T$ is the survey period, $V$ is the comoving volume, $z_i$ is the redshift of an event, and $\varepsilon_i$ is the detection efficiency of an event \citep[e.g.,][]{prajs2017z1slsnrate}. We assume $\varepsilon_i\sim 1$ here for simplicity. The actual efficiency is clearly less than 1 because we could not confirm the redshifts of many high-redshift SLSN candidates and this makes our rates only approximate. The survey period is set as $T\sim 0.5~\mathrm{yr}$. Assuming a typical limiting magnitude of 26.5~mag, SNe evolving with absolute magnitudes brighter than $\simeq -20~\mathrm{mag}$ can be detected up to $z\sim 4.5$.

\begin{figure}
 \begin{center}
  \includegraphics[width=\columnwidth]{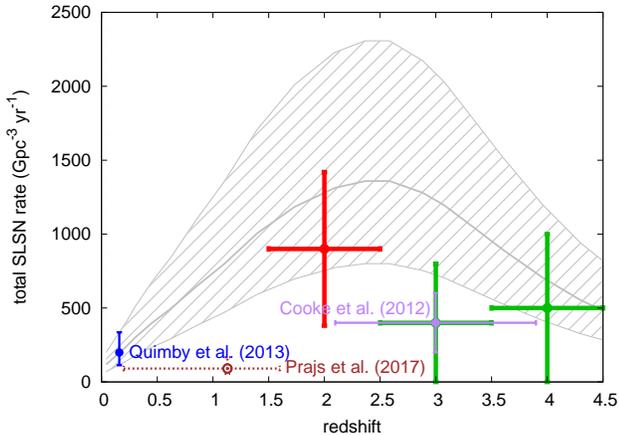}  
 \end{center}
\caption{
Total SLSN rate estimates.
The approximate rates from this study, which are very roughly estimated based on the equation $\sum_{i} (1+z_i)/\varepsilon_i VT$ (see text), are presented with Poisson errors. Our $z\sim2$ rate (red) is based on the spectroscopically confirmed SLSNe and those at $z\sim 3$ and 4 (green) are based on the spectroscopically unconfirmed SLSN candidates. The total SLSN rates estimated in the previous studies \citep{quimby2013slsnrate,cooke2012highzslsn} are also shown. The rate of \citet{cooke2012highzslsn} is a lower limit. We also show the Type~I SLSN rate estimated by \citet{prajs2017z1slsnrate} for reference (dotted brown line, see also \citealt{mccrum2014slsnps111ap}), but the Type~I SLSN rate at $z\sim 0.2$ is only about 10\% of the total SLSN rate \citep{quimby2013slsnrate}. The hatched region shows the total SLSN rates that are extrapolated from the rate at $z\sim 0.16$ \citep{quimby2013slsnrate} based on the cosmic star-formation history \citep{madau2014cosmicsfhis}. We extrapolate the $1\sigma$ range of the rate at $z\sim 0.16$, which is indicated with the hatched region.
}\label{fig:rate}
\end{figure}

We estimate the event rates in the redshift ranges $1.5\lesssim z \lesssim 2.5$, $2.5\lesssim z \lesssim 3.5$, and $3.5\lesssim z \lesssim 4.5$. We confirmed three SLSNe at $1.5\lesssim z \lesssim 2.5$. Summing up the three events, we find the approximate event rate of $\sim 900\pm 520~\mathrm{Gpc^{-3}~yr^{-1}}$ using Poisson error. This rate is consistent with the expected total SLSN rate at $1.5\lesssim z \lesssim 2.5$ obtained by extrapolating the total SLSN rate at $z\sim 0.2$ based on the cosmic star-formation history (Fig.~\ref{fig:rate}).

We obtained at least one SLSN candidate at $2.5\lesssim z \lesssim 3.5$ and another SLSN candidate at $3.5\lesssim z \lesssim 4.5$. The approximate total SLSN rate based on these candidates are $\sim 400\pm400~\mathrm{Gpc^{-3}~yr^{-1}}$ ($2.5\lesssim z \lesssim 3.5$) and $\sim 500\pm500~\mathrm{Gpc^{-3}~yr^{-1}}$ ($3.5\lesssim z \lesssim 4.5$) including the Poisson error. The approximate rates are already consistent with the high-redshift SLSN rate estimated by \citet{cooke2012highzslsn} (Fig.~\ref{fig:rate}). We note that \citet{cooke2012highzslsn} argue that their rate estimate is a lower limit.

The approximate event rates estimated in this section is summarized in Fig.~\ref{fig:rate}. We repeat that the estimates for the event rates presented here are still very approximate. A complete study of the high-redshift SN rates from the HSC-SSP transient survey will be presented elsewhere after the completion of the survey.

\subsection{Host galaxy properties}\label{sec:hostgalaxies}
The host galaxies of the high-redshift SNe and SN candidates reported here have $\sim -21$~mag in the rest ultraviolet and they are relatively massive galaxies ($\sim 10^{10}~\Msun$).
Meanwhile, high-redshift ($z>1$) SLSNe have been classified as Type~I SLSNe and 
typically found in less massive galaxies $(\sim 10^{9-8}~\Msun)$ \citep[e.g.,][]{schulze2016slsnhost}.
The SFRs of our host galaxies ($\sim 10~\Msun~\mathrm{yr^{-1}}$) are also higher than those of the other SLSN host galaxies ($\sim 1~\Msun~\mathrm{yr^{-1}}$) but their specific SFRs ($\sim 10^{-9}~\mathrm{yr^{-1}}$) are similar to those found in the other SLSN host galaxies \citep[e.g.,][]{schulze2016slsnhost}.
The \texttt{MIZUKI} code is shown to give galaxy property estimates that are consistent with the other methods and this difference is not likely to originate from systematic uncertainties \citep{tanaka2015mizuki}.

Previous high-redshift SLSN discoveries are limited to Type~I SLSNe which are found to preferentially appear in low-metallicity environments \citep[e.g.,][]{howell2013slsn,pan2017slsnz1p861,smith2017z2des}.
However, the galaxy mass-metallicity relation at $z\simeq 2$ shows that galaxies with $\sim 10^{10}~\Msun$ at $z\simeq 2$ are as metal-poor as the $\sim 10^{8}-10^9~\Msun$ hosts in the local Universe \citep[e.g.,][]{erb2006massmetallicity,steidel2014massmetallicity,sanders2015massmetallicity}, where Type~I SLSNe are found \citep[e.g.,][]{chen2017toogoodrelation}.
Indeed, high-redshift Type~I SLSNe are sometimes found in such massive galaxies with $\sim 10^{10}~\Msun$ \citep{cooke2012highzslsn,schulze2016slsnhost,perley2016slsnhost}. Type~II SLSN host galaxies have large diversities \citep[e.g.,][]{neill2011slsnhost} and our high-redshift SLSN host properties are consistent with being both Type~II and Type~I host galaxies.

We note, however, that our preference to find SLSNe in relatively massive galaxies could be partly due to our method of selecting high-redshift SN candidates, i.e., using photometric redshifts of the host galaxies. Brighter galaxies would have better estimates of photometric redshifts and we tend to follow high-redshift SN candidates with better photometric redshifts to secure our high-redshift SN discoveries. Further systematic investigations are required to conclude the host galaxy properties of the high-redshift SLSNe.

\section{Conclusions}\label{sec:conclusions}
We have reported our initial high-redshift SN discoveries from the SHIZUCA conducted from November 2016 to May 2017 at the COSMOS field. We have shown three SNe at the redshifts around 2, i.e., HSC16adga at $z=2.399\pm 0.004$, HSC17auzg at $z=1.965\pm 0.004$, and HSC17dbpf at $z=1.851\pm 0.004$. In this paper, we reported their photometric properties and their spectroscopic observations are presented in the accompanying paper \citepalias{curtin2017hscspec}. These high-redshift SNe are first selected based on their host galaxies' photometric redshifts and then the redshifts are confirmed by spectroscopy. They are likely SLSNe because of their proximity to $-21~\mathrm{mag}$ in the rest frame ultraviolet bands that we observed. At least three SLSNe detections at $z\simeq 2$ in our survey using only the spectroscopically confirmed events indicates an approximate SLSN event rate of $\sim 900\pm 520~\mathrm{Gpc^{-3}~yr^{-1}}$ at $z\simeq 2$ including the Poisson error. 

In addition to the spectroscopically confirmed SNe at $z\simeq 2$, we reported one SN candidate at $z\sim 3$ and another SN candidate at $z\sim 4$. Their redshifts are estimated by the photometric redshifts of their host galaxies. The photometric redshifts are based on the photometry of the about 30 bands available in the COSMOS field as well as our HSC photometry. Estimated approximate event rates from the detections are $\sim 400\pm400~\mathrm{Gpc^{-3}~yr^{-1}}$ ($z\sim 3$) and $\sim 500\pm500~\mathrm{Gpc^{-3}~yr^{-1}}$ ($z\sim 4$) with the Poisson errors. 

The high-redshift SNe and SN candidates reported in this paper are still based on our initial analysis, but we can already see the amazing capability of Subaru/HSC to find high-redshift SNe. We have not finished the complete search for high-redshift SNe in our data. In addition, a further transient survey is planned at the SXDS field in 2019--2020. We will obtain more high-redshift SNe from these observations. The complete samples of high-redshift SNe from the HSC-SSP transient survey will be presented after the completion of the survey.

\acknowledgments
This research is supported by the Grants-in-Aid for Scientific Research of the Japan Society for the Promotion of Science (16H07413, 17H02864) and by JSPS Open Partnership Bilateral Joint Research Project between Japan and Chile.
J.C. acknowledges the Australian Research Council Future Fellowship grant FT130101219. 
L.G. was supported in part by the US National Science Foundation under Grant AST-1311862.
Support for G.P. is provided by the Ministry of Economy, Development, and Tourism's Millennium Science Initiative through grant IC120009, awarded to The Millennium Institute of Astrophysics, MAS.
G.P. also acknowledges support by the Proyecto Regular FONDECYT 1140352.
This research is partly supported by Japan Science and Technology Agency CREST JPMHCR1414.
Parts of this research were conducted by the Australian Research Council Centre of Excellence for All-sky Astrophysics (CAASTRO), through project number CE110001020.

The Hyper Suprime-Cam (HSC) collaboration includes the astronomical communities of Japan and Taiwan, and Princeton University.  The HSC instrumentation and software were developed by the National Astronomical Observatory of Japan (NAOJ), the Kavli Institute for the Physics and Mathematics of the Universe (Kavli IPMU), the University of Tokyo, the High Energy Accelerator Research Organization (KEK), the Academia Sinica Institute for Astronomy and Astrophysics in Taiwan (ASIAA), and Princeton University.  Funding was contributed by the FIRST program from Japanese Cabinet Office, the Ministry of Education, Culture, Sports, Science and Technology (MEXT), the Japan Society for the Promotion of Science (JSPS),  Japan Science and Technology Agency  (JST),  the Toray Science  Foundation, NAOJ, Kavli IPMU, KEK, ASIAA,  and Princeton University.

The Pan-STARRS1 Surveys (PS1) have been made possible through contributions of the Institute for Astronomy, the University of Hawaii, the Pan-STARRS Project Office, the Max-Planck Society and its participating institutes, the Max Planck Institute for Astronomy, Heidelberg and the Max Planck Institute for Extraterrestrial Physics, Garching, The Johns Hopkins University, Durham University, the University of Edinburgh, Queen's University Belfast, the Harvard-Smithsonian Center for Astrophysics, the Las Cumbres Observatory Global Telescope Network Incorporated, the National Central University of Taiwan, the Space Telescope Science Institute, the National Aeronautics and Space Administration under Grant No. NNX08AR22G issued through the Planetary Science Division of the NASA Science Mission Directorate, the National Science Foundation under Grant No. AST-1238877, the University of Maryland, and Eotvos Lorand University (ELTE).
 
This paper makes use of software developed for the Large Synoptic Survey Telescope. We thank the LSST Project for making their code available as free software at \url{http://dm.lsst.org}.
  
Based on data collected at the Subaru Telescope and retrieved from the HSC data archive system, which is operated by the Subaru Telescope and Astronomy Data Center at National Astronomical Observatory of Japan.

Based in part on data obtained at the Gemini Observatory via the time exchange program between Gemini and the Subaru Telescope processed using the Gemini IRAF package (program ID: S17A-056, GS-2017A-Q-13). The Gemini Observatory is operated by the Association of Universities for Research in Astronomy, Inc., under a cooperative agreement with the NSF on behalf of the Gemini partnership: the National Science Foundation (United States), the National Research Council (Canada), CONICYT (Chile), Ministerio de Ciencia, Tecnolog\'{i}a e Innovaci\'{o}n Productiva (Argentina), and Minist\'{e}rio da Ci\^{e}ncia, Tecnologia e Inova\c{c}\~{a}o (Brazil).

This research has made use of the NASA/IPAC Extragalactic Database (NED) which is operated by the Jet Propulsion Laboratory, California Institute of Technology, under contract with the National Aeronautics and Space Administration.

\facilities{Subaru/HSC, Gemini/GMOS-S} 
\software{MIZUKI \citep{tanaka2015mizuki}, hscPipe \citep{bosch2017hscpipe}}

\bibliographystyle{yahapj}
\bibliography{references}

\appendix
\section{Log of photometry}
We summarize photometric information of the transients presented in this paper. One count in flux corresponds to 27~mag.
For HSC17dsid, we also show all the image cutouts from HSC because the images in Fig.~\ref{fig:photozfaces} may not be clear enough to identify the detection. The image size is 8.6" x 8.6" (51 pixel x 51 pixel).

\startlongtable
\begin{deluxetable*}{cccccc}
\tablecaption{HSC16adga (SN 2016jhm) \label{tab:hsc16adgaphotometory}}
\tablecolumns{6}
\tablenum{2}
\tablewidth{0pt}
\tablehead{
\colhead{Band} & \colhead{MJD} &
\colhead{Flux} & \colhead{$1\sigma$ flux error} &
\colhead{AB mag.} & \colhead{$1\sigma$ mag. error}  
}
\startdata
$g$&57717.56 &   4.286 &   1.281 & > 24.984 & \nodata \\
&57755.61 &   4.655 &   0.185 & 25.330 &  0.043 \\
&57778.45 &   3.097 &   0.266 & 25.773 &  0.093 \\
&57785.39 &   2.198 &   0.327 & 26.145 &  0.162 \\
&57807.37 &   1.129 &   0.352 & > 26.386 & \nodata \\
&57834.31 &   0.983 &   0.191 & 27.019 &  0.211 \\
&57841.29 &   0.892 &   0.327 & > 26.466 & \nodata \\
&57869.33 &  -0.426 &   0.422 & > 26.189 & \nodata \\
\hline
$r$&57720.60 &   4.554 &   0.367 & 25.354 &  0.087 \\
&57747.53 &   6.734 &   0.390 & 24.929 &  0.063 \\
&57776.40 &   5.092 &   0.261 & 25.233 &  0.056 \\
&57786.45 &   4.099 &   0.319 & 25.468 &  0.084 \\
&57807.48 &   3.260 &   0.318 & 25.717 &  0.106 \\
&57818.52 &   1.471 &   0.265 & 26.581 &  0.195 \\
&57837.26 &   1.174 &   0.403 & > 26.239 & \nodata \\
&57844.33 &   1.807 &   0.422 & > 26.189 & \nodata \\
&57866.25 &   0.655 &   0.746 & > 25.571 & \nodata \\
\hline
$i$&57717.62 &   4.380 &   0.632 & 25.396 &  0.157 \\
&57721.54 &   3.871 &   0.612 & 25.530 &  0.172 \\
&57747.61 &   7.520 &   0.551 & 24.810 &  0.080 \\
&57755.53 &   7.327 &   0.345 & 24.838 &  0.051 \\
&57776.54 &   6.709 &   0.303 & 24.933 &  0.049 \\
&57783.43 &   6.925 &   0.433 & 24.899 &  0.068 \\
&57786.59 &   5.664 &   0.407 & 25.117 &  0.078 \\
&57809.40 &   4.862 &   0.713 & 25.283 &  0.159 \\
&57816.47 &   3.119 &   0.327 & 25.765 &  0.114 \\
&57835.26 &   2.184 &   0.414 & 26.152 &  0.206 \\
&57842.27 &   1.609 &   0.540 & > 25.922 & \nodata \\
&57869.26 &   1.021 &   0.659 & > 25.705 & \nodata \\
&57870.35 &   2.062 &   0.437 & > 26.151 & \nodata \\
\hline
$z$&57715.54 &   0.251 &   1.150 & > 25.101 & \nodata \\
&57721.59 &   4.698 &   0.960 & > 25.297 & \nodata \\
&57745.56 &   8.227 &   0.644 & 24.712 &  0.085 \\
&57755.45 &  10.688 &   0.902 & 24.428 &  0.092 \\
&57774.50 &   8.625 &   0.380 & 24.661 &  0.048 \\
&57779.53 &   9.211 &   1.600 & 24.589 &  0.189 \\
&57783.56 &   7.309 &   0.414 & 24.840 &  0.062 \\
&57805.37 &   6.359 &   0.587 & 24.992 &  0.100 \\
&57816.30 &   5.014 &   0.821 & 25.250 &  0.178 \\
&57834.43 &   5.865 &   0.734 & 25.079 &  0.136 \\
&57841.41 &   3.088 &   0.838 & > 25.444 & \nodata \\
&57866.36 &   3.020 &   1.871 & > 24.572 & \nodata \\
&57872.26 &   4.808 &   1.274 & > 24.990 & \nodata \\
&57924.28 &   5.025 &   3.407 & > 23.922 & \nodata \\
\hline
$y$&57715.62 &   0.397 &   1.855 & > 24.582 & \nodata \\
&57748.54 &  15.618 &   5.070 & > 23.490 & \nodata \\
&57757.52 &  10.330 &   1.545 & 24.465 &  0.162 \\
&57778.62 &   0.598 &   2.643 & > 24.197 & \nodata \\
&57787.46 &  10.169 &   1.569 & 24.482 &  0.168 \\
&57811.40 &   4.843 &   1.733 & > 24.656 & \nodata \\
&57819.45 &   6.652 &   0.758 & 24.943 &  0.124 \\
&57833.37 &   5.724 &   1.245 & > 25.015 & \nodata \\
&57837.49 & -10.941 &   6.880 & > 23.159 & \nodata \\
&57842.35 &   7.829 &   2.245 & > 24.375 & \nodata \\
&57863.28 &   8.006 &   2.423 & > 24.292 & \nodata \\
\enddata
\tablecomments{Non detections are $5\sigma$ limits.}
\end{deluxetable*}

\startlongtable
\begin{deluxetable*}{cccccc}
\tablecaption{HSC17auzg (SN 2016jhn) \label{tab:hsc17auzgphotometory}}
\tablecolumns{6}
\tablenum{3}
\tablewidth{0pt}
\tablehead{
\colhead{Band} & \colhead{MJD} &
\colhead{Flux} & \colhead{$1\sigma$ flux error} &
\colhead{AB mag.} & \colhead{$1\sigma$ mag. error}  
}
\startdata
$g$&57717.57 &   0.636 &   0.519 & > 25.965 & \nodata \\
&57755.62 &   5.269 &   0.178 & 25.196 &  0.037 \\
&57778.45 &   6.749 &   0.241 & 24.927 &  0.039 \\
&57785.39 &   5.611 &   0.269 & 25.127 &  0.052 \\
&57807.37 &   4.862 &   0.263 & 25.283 &  0.059 \\
&57834.32 &   3.475 &   0.191 & 25.648 &  0.060 \\
&57841.29 &   3.467 &   0.257 & 25.650 &  0.081 \\
&57869.33 &   1.546 &   0.248 & 26.527 &  0.174 \\
\hline
$r$&57720.60 &  -0.342 &   0.202 & > 26.989 & \nodata \\
&57747.53 &   2.389 &   0.377 & 26.054 &  0.171 \\
&57776.41 &   9.246 &   0.217 & 24.585 &  0.025 \\
&57786.45 &   9.127 &   0.235 & 24.599 &  0.028 \\
&57807.48 &   8.712 &   0.261 & 24.650 &  0.032 \\
&57818.52 &   8.686 &   0.235 & 24.653 &  0.029 \\
&57837.26 &   6.485 &   0.318 & 24.970 &  0.053 \\
&57844.33 &   4.864 &   0.286 & 25.283 &  0.064 \\
&57866.25 &   3.003 &   0.355 & 25.806 &  0.128 \\
\hline
$i$&57717.62 &   0.672 &   0.368 & > 26.338 & \nodata \\
&57721.55 &   0.648 &   0.447 & > 26.127 & \nodata \\
&57747.61 &   2.426 &   0.466 & 26.038 &  0.209 \\
&57755.51 &   4.749 &   0.232 & 25.308 &  0.053 \\
&57776.55 &   9.451 &   0.269 & 24.561 &  0.031 \\
&57783.43 &  10.661 &   0.388 & 24.430 &  0.040 \\
&57786.58 &   9.828 &   0.476 & 24.519 &  0.053 \\
&57809.41 &  11.621 &   0.421 & 24.337 &  0.039 \\
&57816.48 &  11.767 &   0.246 & 24.323 &  0.023 \\
&57835.27 &  10.867 &   0.447 & 24.410 &  0.045 \\
&57842.27 &   9.276 &   0.321 & 24.582 &  0.038 \\
&57869.27 &   4.634 &   0.490 & 25.335 &  0.115 \\
&57870.35 &   4.646 &   0.376 & 25.332 &  0.088 \\
\hline
$z$&57715.55 &  -0.893 &   0.605 & > 25.798 & \nodata \\
&57721.60 &   0.513 &   0.589 & > 25.827 & \nodata \\
&57745.57 &   2.618 &   0.579 & > 25.846 & \nodata \\
&57755.45 &   6.968 &   0.602 & 24.892 &  0.094 \\
&57774.52 &  11.399 &   0.270 & 24.358 &  0.026 \\
&57779.52 &  13.194 &   1.935 & 24.199 &  0.159 \\
&57783.56 &  12.050 &   0.474 & 24.298 &  0.043 \\
&57805.36 &  14.520 &   0.514 & 24.095 &  0.038 \\
&57816.31 &  15.976 &   0.482 & 23.991 &  0.033 \\
&57834.43 &  14.081 &   0.410 & 24.128 &  0.032 \\
&57841.41 &  12.330 &   0.559 & 24.273 &  0.049 \\
&57866.36 &  10.028 &   0.716 & 24.497 &  0.077 \\
&57872.26 &   7.896 &   0.724 & 24.756 &  0.099 \\
&57924.28 &   2.283 &   2.563 & > 24.231 & \nodata \\
\hline
$y$&57715.62 &  -3.266 &   0.963 & > 25.294 & \nodata \\
&57748.53 &  -0.606 &   5.343 & > 23.433 & \nodata \\
&57757.51 &   8.846 &   1.197 & 24.633 &  0.147 \\
&57778.62 &  10.340 &   2.520 & > 24.249 & \nodata \\
&57787.47 &  11.019 &   0.919 & 24.395 &  0.091 \\
&57811.39 &  10.573 &   2.091 & 24.440 &  0.215 \\
&57819.47 &  15.617 &   0.635 & 24.016 &  0.044 \\
&57833.38 &  16.761 &   0.989 & 23.939 &  0.064 \\
&57837.49 &  19.342 &   5.434 & > 23.415 & \nodata \\
&57842.35 &  15.230 &   1.214 & 24.043 &  0.087 \\
&57863.28 &   9.082 &   1.420 & 24.605 &  0.170 \\
\enddata
\tablecomments{Non detections are $5\sigma$ limits.}
\end{deluxetable*}

\startlongtable
\begin{deluxetable*}{cccccc}
\tablecaption{HSC17dbpf (SN 2017fei) \label{tab:hsc17dbpfphotometory}}
\tablecolumns{6}
\tablenum{4}
\tablewidth{0pt}
\tablehead{
\colhead{Band} & \colhead{MJD} &
\colhead{Flux} & \colhead{$1\sigma$ flux error} &
\colhead{AB mag.} & \colhead{$1\sigma$ mag. error} 
}
\startdata
$g$&57717.57 &   0.038 &   0.749 & > 25.566 & \nodata \\
&57755.61 &  -0.067 &   0.166 & > 27.202 & \nodata \\
&57778.45 &   0.559 &   0.227 & > 26.863 & \nodata \\
&57785.39 &   0.054 &   0.268 & > 26.682 & \nodata \\
&57807.37 &   0.114 &   0.191 & > 27.050 & \nodata \\
&57834.31 &   6.518 &   0.183 & 24.965 &  0.031 \\
&57841.29 &   2.437 &   0.243 & 26.033 &  0.108 \\
&57869.33 &   0.060 &   0.348 & > 26.399 & \nodata \\
\hline
$r$&57720.60 &  -0.422 &   0.249 & > 26.762 & \nodata \\
&57747.53 &  -0.131 &   0.346 & > 26.405 & \nodata \\
&57776.40 &  -0.137 &   0.218 & > 26.906 & \nodata \\
&57786.45 &  -0.019 &   0.221 & > 26.892 & \nodata \\
&57807.48 &  -0.208 &   0.255 & > 26.736 & \nodata \\
&57818.52 &  22.222 &   0.278 & 23.633 &  0.014 \\
&57837.26 &  12.834 &   0.310 & 24.229 &  0.026 \\
&57844.33 &   6.848 &   0.282 & 24.911 &  0.045 \\
&57866.24 &   0.819 &   0.384 & > 26.292 & \nodata \\
\hline
$i$&57717.62 &  -0.802 &   0.386 & > 26.286 & \nodata \\
&57721.55 &   0.487 &   0.467 & > 26.079 & \nodata \\
&57747.62 &  -0.030 &   0.499 & > 26.007 & \nodata \\
&57755.51 &  -0.843 &   0.286 & > 26.612 & \nodata \\
&57776.55 &  -0.049 &   0.263 & > 26.703 & \nodata \\
&57783.43 &  -0.346 &   0.405 & > 26.234 & \nodata \\
&57786.57 &  -0.154 &   0.573 & > 25.857 & \nodata \\
&57809.41 &  -0.090 &   0.485 & > 26.038 & \nodata \\
&57816.47 &  12.005 &   0.259 & 24.302 &  0.023 \\
&57835.26 &  17.798 &   0.425 & 23.874 &  0.026 \\
&57842.27 &  13.090 &   0.354 & 24.208 &  0.029 \\
&57869.27 &   2.364 &   0.506 & > 25.992 & \nodata \\
&57870.35 &   3.385 &   0.453 & 25.676 &  0.145 \\
\hline
$z$&57715.55 &   0.050 &   0.643 & > 25.732 & \nodata \\
&57721.60 &   0.463 &   0.789 & > 25.510 & \nodata \\
&57745.56 &   0.274 &   0.648 & > 25.724 & \nodata \\
&57755.45 &  -0.961 &   0.752 & > 25.562 & \nodata \\
&57774.50 &  -0.062 &   0.253 & > 26.745 & \nodata \\
&57783.55 &   0.327 &   0.425 & > 26.182 & \nodata \\
&57805.37 &  -0.036 &   0.469 & > 26.075 & \nodata \\
&57816.31 &  12.154 &   0.548 & 24.288 &  0.049 \\
&57834.43 &  18.671 &   0.461 & 23.822 &  0.027 \\
&57841.41 &  16.124 &   0.580 & 23.981 &  0.039 \\
&57866.36 &   5.048 &   0.825 & 25.242 &  0.177 \\
&57872.26 &   4.154 &   0.766 & 25.454 &  0.200 \\
&57924.28 &   3.222 &   2.895 & > 24.098 & \nodata \\
\hline
$y$&57715.62 &  -1.708 &   1.030 & > 25.220 & \nodata \\
&57748.53 &  -7.152 &   5.735 & > 23.356 & \nodata \\
&57757.52 &   3.738 &   1.216 & > 25.040 & \nodata \\
&57778.62 &  -0.099 &   2.208 & > 24.393 & \nodata \\
&57787.48 &   0.118 &   0.914 & > 25.350 & \nodata \\
&57811.40 &  -0.344 &   1.561 & > 24.769 & \nodata \\
&57819.46 &  16.211 &   0.745 & 23.975 &  0.050 \\
&57833.38 &  14.346 &   0.974 & 24.108 &  0.074 \\
&57837.49 &  26.400 &   6.482 & > 23.223 & \nodata \\
&57842.35 &  12.303 &   1.476 & 24.275 &  0.130 \\
&57863.28 &   9.221 &   1.714 & 24.588 &  0.202 \\
\enddata
\tablecomments{Non detections are $5\sigma$ limits.}
\end{deluxetable*}

\startlongtable
\begin{deluxetable*}{cccccc}
\tablecaption{HSC16apuo (AT 2016jho) \label{tab:hsc16apuophotometory}}
\tablecolumns{6}
\tablenum{5}
\tablewidth{0pt}
\tablehead{
\colhead{Band} & \colhead{MJD} &
\colhead{Flux} & \colhead{$1\sigma$ flux error} &
\colhead{AB mag.} & \colhead{$1\sigma$ mag. error} 
}
\startdata
$g$&57717.56 &  -0.277 &   0.582 & > 25.840 & \nodata  \\
&57755.61 &   2.165 &   0.182 & 26.161 &  0.091 \\
&57778.45 &   1.180 &   0.255 & > 26.736 & \nodata \\
&57785.39 &  -0.060 &   0.232 & > 26.839 & \nodata \\
&57807.37 &  -0.392 &   0.211 & > 26.942 & \nodata \\
&57834.32 &  -0.226 &   0.178 & > 27.127 & \nodata \\
&57841.29 &   0.343 &   0.257 & > 26.728 & \nodata \\
&57869.33 &  -0.681 &   0.242 & > 26.793 & \nodata \\
\hline
$r$&57720.60 &  -0.063 &   0.200 & > 27.000 & \nodata \\
&57747.53 &   5.452 &   0.360 & 25.159 &  0.072 \\
&57776.42 &   5.664 &   0.228 & 25.117 &  0.044 \\
&57786.45 &   3.087 &   0.221 & 25.776 &  0.078 \\
&57807.48 &   0.621 &   0.221 & > 26.892 & \nodata \\
&57818.52 &   0.119 &   0.238 & > 26.811 & \nodata \\
&57837.26 &   0.426 &   0.376 & > 26.315 & \nodata \\
&57844.33 &   0.031 &   0.299 & > 26.563 & \nodata \\
&57866.25 &   0.149 &   0.358 & > 26.368 & \nodata \\
\hline
$i$&57717.62 &  -0.770 &   0.381 & > 26.300 & \nodata \\
&57721.55 &   0.006 &   0.466 & > 26.082 & \nodata \\
&57747.62 &   7.054 &   0.425 & 24.879 &  0.065 \\
&57755.51 &  11.355 &   0.303 & 24.362 &  0.029 \\
&57776.55 &  11.516 &   0.271 & 24.347 &  0.026 \\
&57783.43 &   9.073 &   0.344 & 24.606 &  0.041 \\
&57786.59 &   8.552 &   0.468 & 24.670 &  0.059 \\
&57809.41 &   3.356 &   0.414 & 25.685 &  0.134 \\
&57816.47 &   2.330 &   0.225 & 26.081 &  0.105 \\
&57835.26 &   1.747 &   0.487 & > 26.034 & \nodata \\
&57842.27 &   1.172 &   0.366 & > 26.344 & \nodata \\
&57869.26 &   1.004 &   0.604 & > 25.800 & \nodata \\
&57870.35 &   1.014 &   0.363 & > 26.353 & \nodata \\
\hline
$z$&57715.55 &   0.771 &   0.683 & > 25.667 & \nodata \\
&57721.60 &   0.482 &   0.590 & > 25.825 & \nodata \\
&57745.56 &   5.031 &   0.592 & 25.246 &  0.128 \\
&57755.45 &  10.683 &   0.530 & 24.428 &  0.054 \\
&57774.51 &  13.982 &   0.249 & 24.136 &  0.019 \\
&57779.53 &  10.738 &   2.171 & > 24.411 & \nodata \\
&57783.55 &  12.158 &   0.418 & 24.288 &  0.037 \\
&57805.37 &   4.962 &   0.510 & 25.261 &  0.112 \\
&57816.31 &   4.369 &   0.478 & 25.399 &  0.119 \\
&57834.43 &   1.976 &   0.444 & > 26.134 & \nodata \\
&57841.41 &   2.062 &   0.540 & > 25.922 & \nodata \\
&57866.36 &   1.890 &   0.745 & > 25.572 & \nodata \\
&57872.26 &   2.513 &   0.714 & > 25.618 & \nodata \\
&57924.28 &   1.959 &   3.274 & > 23.965 & \nodata \\
\hline
$y$&57715.62 &   0.027 &   0.903 & > 25.363 & \nodata \\
&57748.53 &   7.992 &   5.155 & > 23.472 & \nodata \\
&57757.51 &  12.356 &   1.171 & 24.270 &  0.103 \\
&57778.62 &  14.631 &   1.915 & 24.087 &  0.142 \\
&57787.47 &  13.750 &   0.839 & 24.154 &  0.066 \\
&57811.39 &   8.701 &   1.407 & 24.651 &  0.176 \\
&57819.47 &   5.529 &   0.615 & 25.143 &  0.121 \\
&57833.37 &   3.701 &   1.161 & > 25.090 & \nodata \\
&57837.49 &  16.544 &   5.861 & > 23.333 & \nodata \\
&57842.35 &   1.683 &   1.162 & > 25.090 & \nodata \\
&57863.28 &   6.796 &   1.397 & > 24.890 & \nodata \\
\enddata
\tablecomments{Non detections are $5\sigma$ limits.}
\end{deluxetable*}

\startlongtable
\begin{deluxetable*}{ccccccccc}
\tablecaption{HSC17dsid (AT~2017fej) \label{tab:hsc17dsidphotometory}}
\tablecolumns{9}
\tablenum{6}
\tablewidth{0pt}
\tablehead{
\colhead{Band} & \colhead{MJD} &
\colhead{Flux} & \colhead{$1\sigma$ flux error} &
\colhead{AB mag.} & \colhead{$1\sigma$ mag. error} 
& \colhead{Reference} & \colhead{Observation} & \colhead{Subtraction}
}
\startdata
$g$&57717.57 &  -0.709 &   0.700 & > 25.640 & \nodata & \raisebox{-\totalheight}{\includegraphics[]{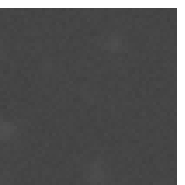}} & \raisebox{-\totalheight}{\includegraphics[]{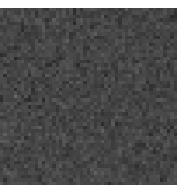}} & \raisebox{-\totalheight}{\includegraphics[]{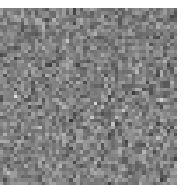}} \\
&57755.62 &   0.198 &   0.216 & > 26.916 & \nodata & \raisebox{-\totalheight}{\includegraphics[]{figures/g_ref.eps}} & \raisebox{-\totalheight}{\includegraphics[]{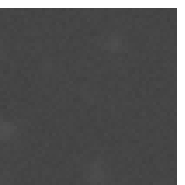}} & \raisebox{-\totalheight}{\includegraphics[]{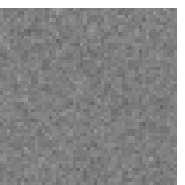}}\\
&57778.45 &  -1.001 &   0.218 & > 26.906 & \nodata & \raisebox{-\totalheight}{\includegraphics[]{figures/g_ref.eps}} & \raisebox{-\totalheight}{\includegraphics[]{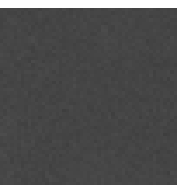}} & \raisebox{-\totalheight}{\includegraphics[]{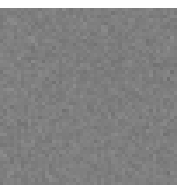}}\\
&57785.39 &   0.378 &   0.223 & > 26.882 & \nodata & \raisebox{-\totalheight}{\includegraphics[]{figures/g_ref.eps}} & \raisebox{-\totalheight}{\includegraphics[]{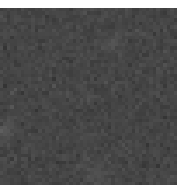}} & \raisebox{-\totalheight}{\includegraphics[]{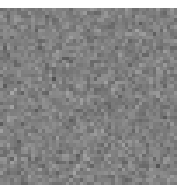}}\\
&57807.36 &  -0.541 &   0.369 & > 26.335 & \nodata & \raisebox{-\totalheight}{\includegraphics[]{figures/g_ref.eps}} & \raisebox{-\totalheight}{\includegraphics[]{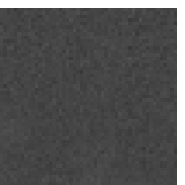}} & \raisebox{-\totalheight}{\includegraphics[]{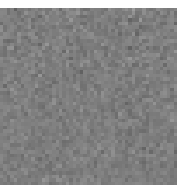}}\\
&57834.32 &  -0.048 &   0.204 & > 26.978 & \nodata & \raisebox{-\totalheight}{\includegraphics[]{figures/g_ref.eps}} & \raisebox{-\totalheight}{\includegraphics[]{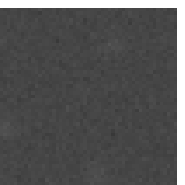}} & \raisebox{-\totalheight}{\includegraphics[]{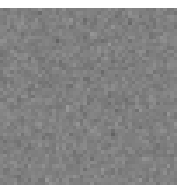}}\\
&57841.29 &   0.221 &   0.224 & > 26.877 & \nodata & \raisebox{-\totalheight}{\includegraphics[]{figures/g_ref.eps}} & \raisebox{-\totalheight}{\includegraphics[]{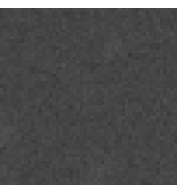}} & \raisebox{-\totalheight}{\includegraphics[]{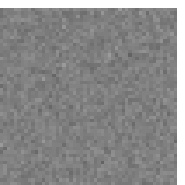}}\\
&57869.33 &   0.898 &   0.278 & > 26.642 & \nodata & \raisebox{-\totalheight}{\includegraphics[]{figures/g_ref.eps}} & \raisebox{-\totalheight}{\includegraphics[]{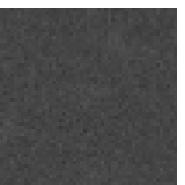}} & \raisebox{-\totalheight}{\includegraphics[]{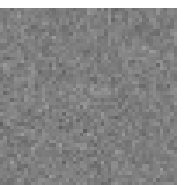}}\\ \\
\hline
$r$&57720.60 &  -0.189 &   0.263 & > 26.703 & \nodata & \raisebox{-\totalheight}{\includegraphics[]{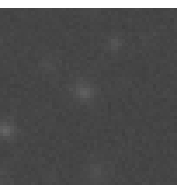}} & \raisebox{-\totalheight}{\includegraphics[]{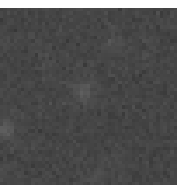}} & \raisebox{-\totalheight}{\includegraphics[]{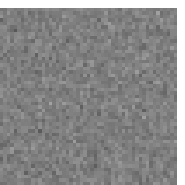}} \\
&57747.53 &  -0.134 &   0.437 & > 26.151 & \nodata & \raisebox{-\totalheight}{\includegraphics[]{figures/r_ref.eps}} & \raisebox{-\totalheight}{\includegraphics[]{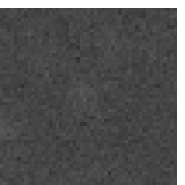}} & \raisebox{-\totalheight}{\includegraphics[]{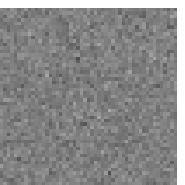}} \\
&57776.40 &  -0.058 &   0.233 & > 26.834 & \nodata & \raisebox{-\totalheight}{\includegraphics[]{figures/r_ref.eps}} & \raisebox{-\totalheight}{\includegraphics[]{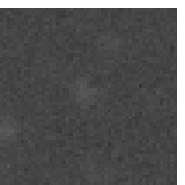}} & \raisebox{-\totalheight}{\includegraphics[]{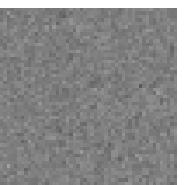}} \\
&57786.45 &   0.650 &   0.252 & > 26.749 & \nodata & \raisebox{-\totalheight}{\includegraphics[]{figures/r_ref.eps}} & \raisebox{-\totalheight}{\includegraphics[]{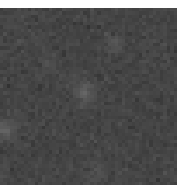}} & \raisebox{-\totalheight}{\includegraphics[]{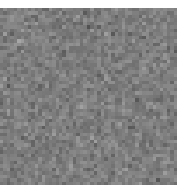}}\\
&57807.48 &  -0.348 &   0.249 & > 26.762 & \nodata & \raisebox{-\totalheight}{\includegraphics[]{figures/r_ref.eps}} & \raisebox{-\totalheight}{\includegraphics[]{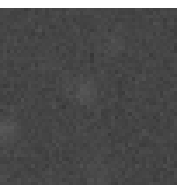}} & \raisebox{-\totalheight}{\includegraphics[]{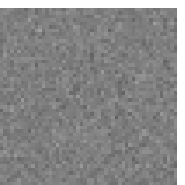}} \\
&57818.51 &  -0.255 &   0.233 & > 26.834 & \nodata & \raisebox{-\totalheight}{\includegraphics[]{figures/r_ref.eps}} & \raisebox{-\totalheight}{\includegraphics[]{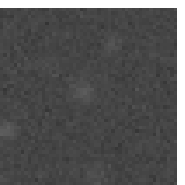}} & \raisebox{-\totalheight}{\includegraphics[]{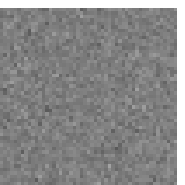}}\\
&57837.26 &   0.445 &   0.337 & > 26.434 & \nodata & \raisebox{-\totalheight}{\includegraphics[]{figures/r_ref.eps}} & \raisebox{-\totalheight}{\includegraphics[]{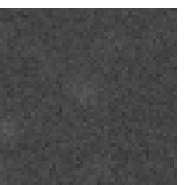}} & \raisebox{-\totalheight}{\includegraphics[]{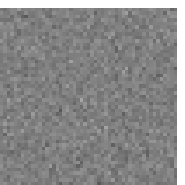}} \\
&57844.34 &  -0.294 &   0.390 & > 26.275 & \nodata & \raisebox{-\totalheight}{\includegraphics[]{figures/r_ref.eps}} & \raisebox{-\totalheight}{\includegraphics[]{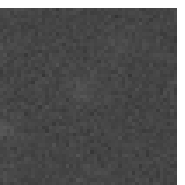}} & \raisebox{-\totalheight}{\includegraphics[]{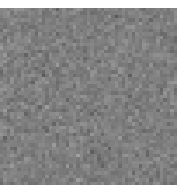}}\\
&57866.25 &   0.788 &   0.330 & > 26.456 & \nodata & \raisebox{-\totalheight}{\includegraphics[]{figures/r_ref.eps}} & \raisebox{-\totalheight}{\includegraphics[]{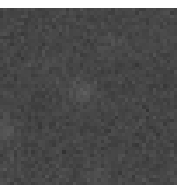}} & \raisebox{-\totalheight}{\includegraphics[]{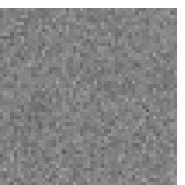}}\\ \\
\hline
$i$&57717.61 &   0.529 &   0.498 & > 26.010 & \nodata & \raisebox{-\totalheight}{\includegraphics[]{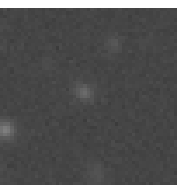}} & \raisebox{-\totalheight}{\includegraphics[]{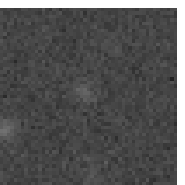}} & \raisebox{-\totalheight}{\includegraphics[]{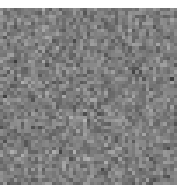}} \\
&57721.55 &  -0.158 &   0.456 & > 26.105 & \nodata & \raisebox{-\totalheight}{\includegraphics[]{figures/i_ref.eps}} & \raisebox{-\totalheight}{\includegraphics[]{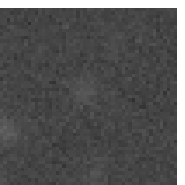}} & \raisebox{-\totalheight}{\includegraphics[]{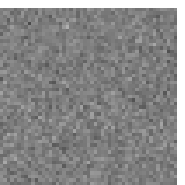}} \\
&57747.62 &  -0.533 &   0.470 & > 26.072 & \nodata & \raisebox{-\totalheight}{\includegraphics[]{figures/i_ref.eps}} & \raisebox{-\totalheight}{\includegraphics[]{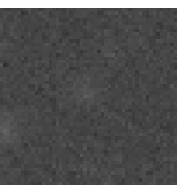}} & \raisebox{-\totalheight}{\includegraphics[]{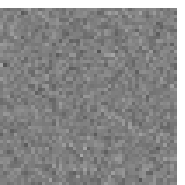}} \\
&57755.52 &  -0.607 &   0.270 & > 26.674 & \nodata & \raisebox{-\totalheight}{\includegraphics[]{figures/i_ref.eps}} & \raisebox{-\totalheight}{\includegraphics[]{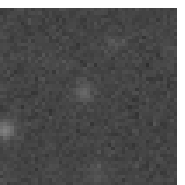}} & \raisebox{-\totalheight}{\includegraphics[]{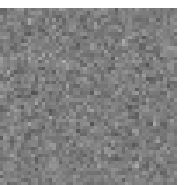}} \\
&57776.54 &  -0.275 &   0.251 & > 26.753 & \nodata  & \raisebox{-\totalheight}{\includegraphics[]{figures/i_ref.eps}} & \raisebox{-\totalheight}{\includegraphics[]{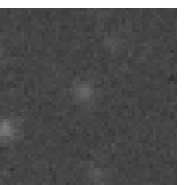}} & \raisebox{-\totalheight}{\includegraphics[]{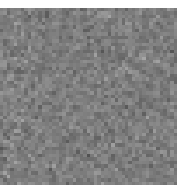}}\\
&57783.43 &   0.021 &   0.356 & > 26.374 & \nodata & \raisebox{-\totalheight}{\includegraphics[]{figures/i_ref.eps}} & \raisebox{-\totalheight}{\includegraphics[]{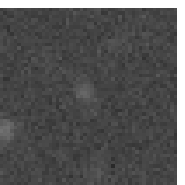}} & \raisebox{-\totalheight}{\includegraphics[]{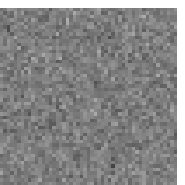}} \\
&57786.58 &  -0.359 &   0.356 & > 26.374 & \nodata & \raisebox{-\totalheight}{\includegraphics[]{figures/i_ref.eps}} & \raisebox{-\totalheight}{\includegraphics[]{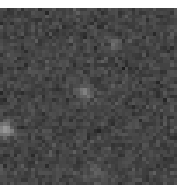}} & \raisebox{-\totalheight}{\includegraphics[]{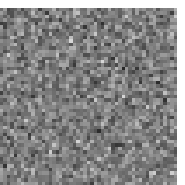}} \\
&57809.41 &  -0.133 &   0.455 & > 26.108 & \nodata & \raisebox{-\totalheight}{\includegraphics[]{figures/i_ref.eps}} & \raisebox{-\totalheight}{\includegraphics[]{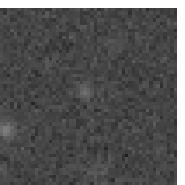}} & \raisebox{-\totalheight}{\includegraphics[]{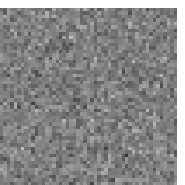}} \\
&57816.47 &  -0.127 &   0.282 & > 26.627 & \nodata & \raisebox{-\totalheight}{\includegraphics[]{figures/i_ref.eps}} & \raisebox{-\totalheight}{\includegraphics[]{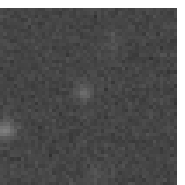}} & \raisebox{-\totalheight}{\includegraphics[]{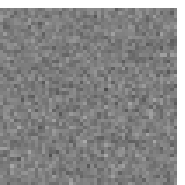}} \\
&57835.26 &  -0.216 &   0.394 & > 26.264 & \nodata & \raisebox{-\totalheight}{\includegraphics[]{figures/i_ref.eps}} & \raisebox{-\totalheight}{\includegraphics[]{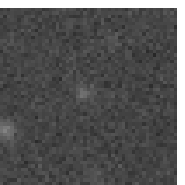}} & \raisebox{-\totalheight}{\includegraphics[]{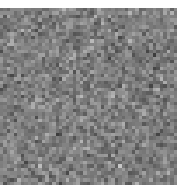}} \\
&57842.27 &  -0.209 &   0.396 & > 26.258 & \nodata & \raisebox{-\totalheight}{\includegraphics[]{figures/i_ref.eps}} & \raisebox{-\totalheight}{\includegraphics[]{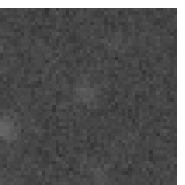}} & \raisebox{-\totalheight}{\includegraphics[]{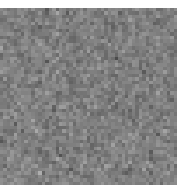}} \\
&57869.26 &   3.999 &   0.629 & 25.495 &  0.171 & \raisebox{-\totalheight}{\includegraphics[]{figures/i_ref.eps}} & \raisebox{-\totalheight}{\includegraphics[]{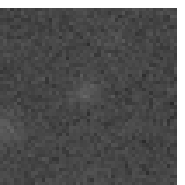}} & \raisebox{-\totalheight}{\includegraphics[]{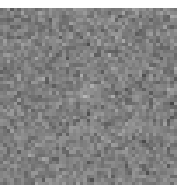}} \\
&57870.35 &   4.277 &   0.410 & 25.422 &  0.104  & \raisebox{-\totalheight}{\includegraphics[]{figures/i_ref.eps}} & \raisebox{-\totalheight}{\includegraphics[]{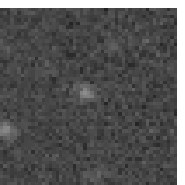}} & \raisebox{-\totalheight}{\includegraphics[]{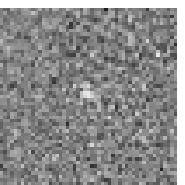}} \\
&57904.98\tablenotemark{a} & 43.65\tablenotemark{a} & 2.04\tablenotemark{a} & 22.9\tablenotemark{a} &0.05\tablenotemark{a} & \nodata\tablenotemark{a} & \nodata\tablenotemark{a} & \nodata\tablenotemark{a} \\
\hline
$z$&57715.55 &   1.349 &   0.716 & > 25.615 & \nodata & \raisebox{-\totalheight}{\includegraphics[]{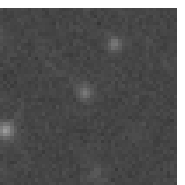}} & \raisebox{-\totalheight}{\includegraphics[]{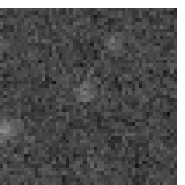}} & \raisebox{-\totalheight}{\includegraphics[]{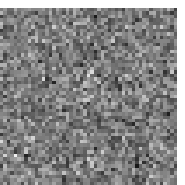}} \\
&57721.60 &  -2.098 &   0.596 & > 25.814 & \nodata & \raisebox{-\totalheight}{\includegraphics[]{figures/z_ref.eps}} & \raisebox{-\totalheight}{\includegraphics[]{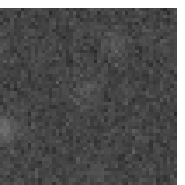}} & \raisebox{-\totalheight}{\includegraphics[]{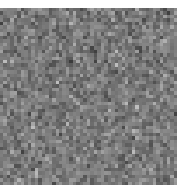}} \\
&57745.57 &  -1.133 &   0.646 & > 25.727 & \nodata & \raisebox{-\totalheight}{\includegraphics[]{figures/z_ref.eps}} & \raisebox{-\totalheight}{\includegraphics[]{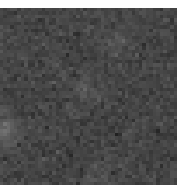}} & \raisebox{-\totalheight}{\includegraphics[]{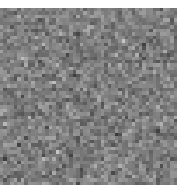}} \\
&57755.45 &   0.684 &   0.624 & > 25.765 & \nodata & \raisebox{-\totalheight}{\includegraphics[]{figures/z_ref.eps}} & \raisebox{-\totalheight}{\includegraphics[]{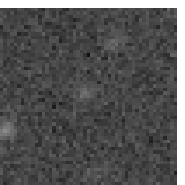}} & \raisebox{-\totalheight}{\includegraphics[]{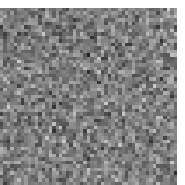}} \\
&57774.50 &   0.456 &   0.252 & > 26.749 & \nodata & \raisebox{-\totalheight}{\includegraphics[]{figures/z_ref.eps}} & \raisebox{-\totalheight}{\includegraphics[]{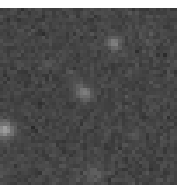}} & \raisebox{-\totalheight}{\includegraphics[]{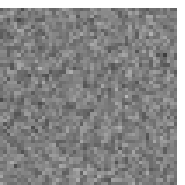}} \\
&57779.52 &  -1.118 &   1.451 & > 24.848 & \nodata & \raisebox{-\totalheight}{\includegraphics[]{figures/z_ref.eps}} & \raisebox{-\totalheight}{\includegraphics[]{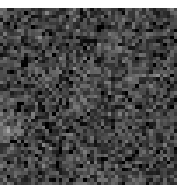}} & \raisebox{-\totalheight}{\includegraphics[]{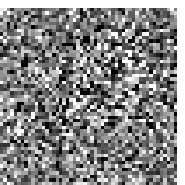}} \\
&57783.55 &  -1.195 &   0.486 & > 26.036 & \nodata & \raisebox{-\totalheight}{\includegraphics[]{figures/z_ref.eps}} & \raisebox{-\totalheight}{\includegraphics[]{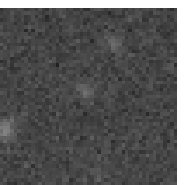}} & \raisebox{-\totalheight}{\includegraphics[]{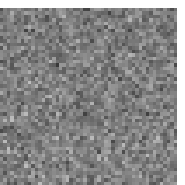}} \\
&57805.37 &   0.101 &   0.547 & > 25.908 & \nodata & \raisebox{-\totalheight}{\includegraphics[]{figures/z_ref.eps}} & \raisebox{-\totalheight}{\includegraphics[]{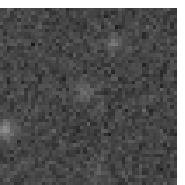}} & \raisebox{-\totalheight}{\includegraphics[]{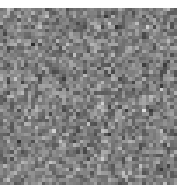}} \\
&57816.31 &  -1.782 &   0.528 & > 25.946 & \nodata & \raisebox{-\totalheight}{\includegraphics[]{figures/z_ref.eps}} & \raisebox{-\totalheight}{\includegraphics[]{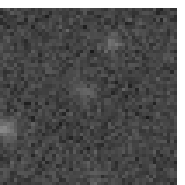}} & \raisebox{-\totalheight}{\includegraphics[]{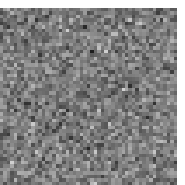}} \\
&57834.44 &   0.352 &   0.486 & > 26.036 & \nodata & \raisebox{-\totalheight}{\includegraphics[]{figures/z_ref.eps}} & \raisebox{-\totalheight}{\includegraphics[]{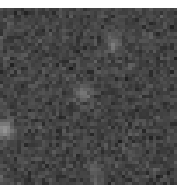}} & \raisebox{-\totalheight}{\includegraphics[]{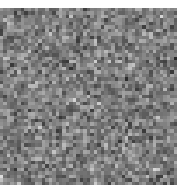}} \\
&57841.41 &  -1.159 &   0.766 & > 25.542 & \nodata & \raisebox{-\totalheight}{\includegraphics[]{figures/z_ref.eps}} & \raisebox{-\totalheight}{\includegraphics[]{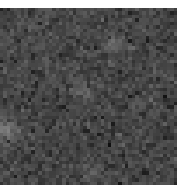}} & \raisebox{-\totalheight}{\includegraphics[]{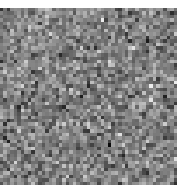}} \\
&57866.36 &  -0.539 &   0.872 & > 25.401 & \nodata & \raisebox{-\totalheight}{\includegraphics[]{figures/z_ref.eps}} & \raisebox{-\totalheight}{\includegraphics[]{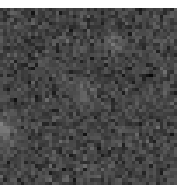}} & \raisebox{-\totalheight}{\includegraphics[]{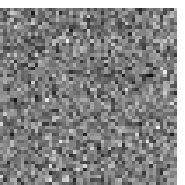}} \\
&57872.26 &   6.609 &   0.880 & 24.950 &  0.145 & \raisebox{-\totalheight}{\includegraphics[]{figures/z_ref.eps}} & \raisebox{-\totalheight}{\includegraphics[]{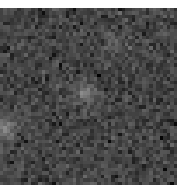}} & \raisebox{-\totalheight}{\includegraphics[]{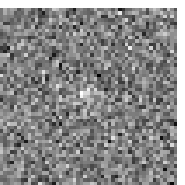}} \\
&57924.28 &  18.409 &   2.630 & 23.837 &  0.155 & \raisebox{-\totalheight}{\includegraphics[]{figures/z_ref.eps}} & \raisebox{-\totalheight}{\includegraphics[]{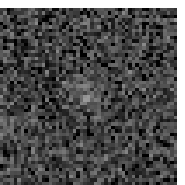}} & \raisebox{-\totalheight}{\includegraphics[]{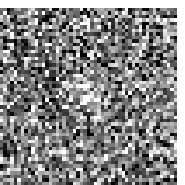}} \\ \\
\hline
$y$&57715.62 &  -1.774 &   0.882 & > 25.389 & \nodata & \raisebox{-\totalheight}{\includegraphics[]{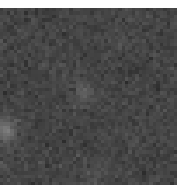}} & \raisebox{-\totalheight}{\includegraphics[]{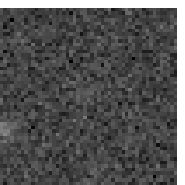}} & \raisebox{-\totalheight}{\includegraphics[]{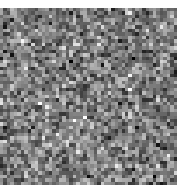}} \\
&57748.54 &  -4.335 &   5.787 & > 23.346 & \nodata & \raisebox{-\totalheight}{\includegraphics[]{figures/y_ref.eps}} & \raisebox{-\totalheight}{\includegraphics[]{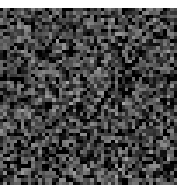}} & \raisebox{-\totalheight}{\includegraphics[]{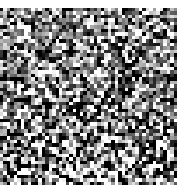}} \\
&57757.53 &  -1.652 &   1.199 & > 25.056 & \nodata & \raisebox{-\totalheight}{\includegraphics[]{figures/y_ref.eps}} & \raisebox{-\totalheight}{\includegraphics[]{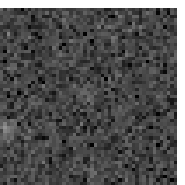}} & \raisebox{-\totalheight}{\includegraphics[]{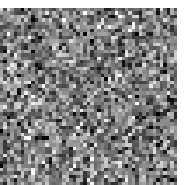}} \\
&57778.62 &  -0.543 &   2.108 & > 24.443 & \nodata & \raisebox{-\totalheight}{\includegraphics[]{figures/y_ref.eps}} & \raisebox{-\totalheight}{\includegraphics[]{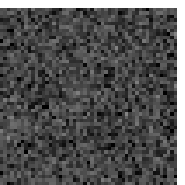}} & \raisebox{-\totalheight}{\includegraphics[]{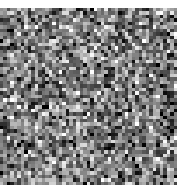}} \\
&57787.47 &  -0.722 &   0.818 & > 25.471 & \nodata & \raisebox{-\totalheight}{\includegraphics[]{figures/y_ref.eps}} & \raisebox{-\totalheight}{\includegraphics[]{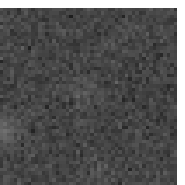}} & \raisebox{-\totalheight}{\includegraphics[]{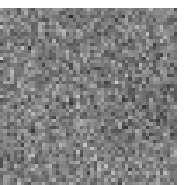}} \\
&57811.40 &  -1.233 &   1.449 & > 24.850 & \nodata & \raisebox{-\totalheight}{\includegraphics[]{figures/y_ref.eps}} & \raisebox{-\totalheight}{\includegraphics[]{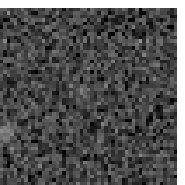}} & \raisebox{-\totalheight}{\includegraphics[]{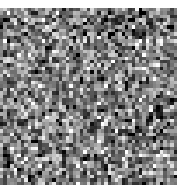}} \\
&57819.47 &  -1.443 &   0.714 & > 25.618 & \nodata & \raisebox{-\totalheight}{\includegraphics[]{figures/y_ref.eps}} & \raisebox{-\totalheight}{\includegraphics[]{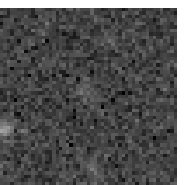}} & \raisebox{-\totalheight}{\includegraphics[]{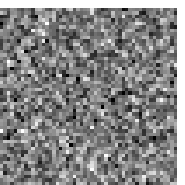}} \\
&57833.38 &  -0.814 &   1.148 & > 25.103 & \nodata & \raisebox{-\totalheight}{\includegraphics[]{figures/y_ref.eps}} & \raisebox{-\totalheight}{\includegraphics[]{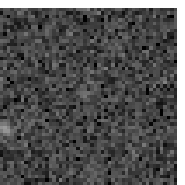}} & \raisebox{-\totalheight}{\includegraphics[]{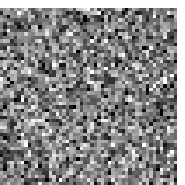}} \\
&57837.49 &   8.798 &   5.487 & > 23.404 & \nodata & \raisebox{-\totalheight}{\includegraphics[]{figures/y_ref.eps}} & \raisebox{-\totalheight}{\includegraphics[]{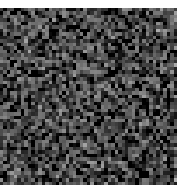}} & \raisebox{-\totalheight}{\includegraphics[]{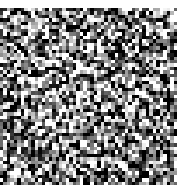}} \\
&57842.35 &  -0.071 &   1.262 & > 25.000 & \nodata & \raisebox{-\totalheight}{\includegraphics[]{figures/y_ref.eps}} & \raisebox{-\totalheight}{\includegraphics[]{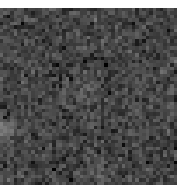}} & \raisebox{-\totalheight}{\includegraphics[]{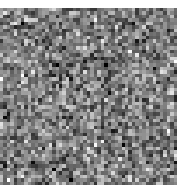}} \\
&57863.28 &  -1.771 &   1.716 & > 24.666 & \nodata & \raisebox{-\totalheight}{\includegraphics[]{figures/y_ref.eps}} & \raisebox{-\totalheight}{\includegraphics[]{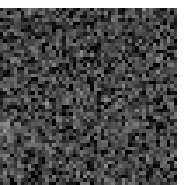}} & \raisebox{-\totalheight}{\includegraphics[]{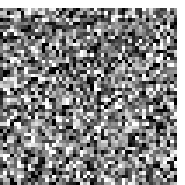}} \\ \\
\enddata
\tablenotetext{a}{Observed with Gemini/GMOS-S.}
\tablecomments{Non detections are $5\sigma$ limits.}
\end{deluxetable*}

\textcolor{white}{0}


\end{document}